\documentclass[10pt,journal,compsoc]{IEEEtran}
\usepackage{booktabs} 

\usepackage[normalem]{ulem}
\usepackage{graphicx,dblfloatfix}
\usepackage{amsmath}
\usepackage{soul}
\usepackage[linesnumbered,ruled]{algorithm2e}
\usepackage{subfig}
\usepackage{relsize}
\usepackage{multirow}
\usepackage{hhline}
\usepackage{dirtytalk}
\usepackage{xcolor}
\usepackage{soul}
\usepackage{longtable}
\usepackage{tabularx}
\usepackage{afterpage}
\usepackage{rotating}
\usepackage{perpage}
\usepackage{float}
\usepackage{mathtools}
\usepackage{enumitem}
\usepackage{dblfloatfix}
\usepackage{ragged2e} 
\usepackage{tikz}
\usepackage{xcolor}

\usepackage{rotating}
\usepackage{bbding}
\usepackage{chngpage} 

\newcommand*\circled[1]{\tikz[baseline=(char.base)]{
		\node[shape=circle,fill,inner sep=0.7pt] (char) {\textcolor{white}{#1}};}}

 \definecolor{mytxtcolor}{rgb}{1.0, 0.01, 0.24}

\SetCommentSty{mycommfont}

\newcommand\techname{ETICA}

\newcommand\shfullname{\underline{E}fficient \underline{T}wo-Level \underline{I}/O \underline{C}aching \underline{A}rchitecture}

\newcommand\URDP{POD}
\newcommand\URDPfull{Policy Optimized reuse Distance}

\newcommand{\wdev}{wdev\_0}

\newcommand{\webse}{web\_3}
\newcommand{\stg}{stg\_1}
\newcommand{\ts}{ts\_0}
\newcommand{\hm}{hm\_1}
\newcommand{\mdssefr}{mds\_0}
\newcommand{\proj}{proj\_0}

\newcommand{\rsrchsefr}{rsrch\_0}
\newcommand{\srcyek}{src1\_2}

\newcommand{\srcdo}{src2\_0}

\newcommand{\usr}{usr\_0}

\newcommand{\mdsyek}{mds\_1}
\ifCLASSOPTIONcompsoc
  \usepackage[nocompress]{cite}
\else
  \usepackage{cite}
\fi
\ifCLASSINFOpdf
\else
\fi
\begin{document}
%
\title{\techname{}: \underline{E}fficient \underline{T}wo-Level \underline{I}/O \underline{C}aching \underline{A}rchitecture for Virtualized Platforms}

	\author{Saba~Ahmadian,~Reza~Salkhordeh,~Onur~Mutlu,~Hossein~Asadi}

\IEEEtitleabstractindextext{%
	{
		\justify
\begin{abstract}
	In recent years, 
	increased I/O demand of \emph{Virtual Machines} (VMs) in large-scale data centers and cloud {computing} has encouraged system architects to design high-performance storage systems.
	One common approach to improving performance is to employ fast storage devices such as \emph{Solid-State Drives} (SSDs) as {an I/O} caching layer for {slower} storage {devices}.
	SSDs provide high performance, especially on random {requests,} but {they {also} have limited endurance: they support \emph{only} a limited number of  write operations and can therefore \emph{wear out} relatively fast due to write operations.}
	In addition to the write requests {generated by the applications}, each read miss in {the} {SSD} cache is served at the cost of imposing a write operation to the SSD (to copy the data block into the cache), resulting in {an even} {larger} number of writes {into the SSD}.
	Previous I/O caching schemes on virtualized platforms \emph{only} partially mitigate the endurance limitations of SSD-based I/O {caches;} they mainly focus on assigning efficient cache write {policies} and cache {space} to the VMs.
	Moreover, existing cache space allocation schemes {have inefficiencies: they} \emph{do not} take into account the impact of cache write policy in reuse distance calculation of the running workloads and hence, reserve cache blocks for accesses {that} would \emph{not} be served by cache.

	{In this paper, we propose an \emph{\shfullname{}} (\techname{}) for virtualized platforms {that} can significantly improve I/O latency, endurance, and cost {(in terms of cache size)} while preserving the reliability of write-pending data blocks.}
	As opposed to previous \emph{one-level} I/O caching schemes in virtualized platforms, our proposed architecture {1)} provides \emph{two} levels of cache by 
	employing {both} \emph{Dynamic Random-Access Memory} (DRAM) and SSD in the I/O caching layer of virtualized platforms and {2)} effectively partitions the cache space between running VMs to achieve maximum performance and minimum {cache size}.
	{To manage the two-level cache,} unlike the previous reuse distance calculation schemes such as \emph{Useful Reuse Distance} (URD), which only consider the request type and neglect the impact of
	\emph{cache write policy}, we propose a new metric, \emph{\URDPfull{}} (\URDP{}).
	{The key idea of \URDP{} is to} effectively {calculate} the reuse distance and {estimate} the {amount of two-level DRAM+SSD} cache space {to allocate} by considering both 1) {the} request type and 2) {the} cache write policy. Doing so results in enhanced performance and reduced {cache size} {due to the}  {allocation of} cache blocks {\emph{only}} for the requests {that} would be {served by the} I/O cache.
	{\techname{} maintains the reliability {of write-pending data blocks} and improves performance by 1) assigning an {effective} and fixed write policy at each level of {the} I/O cache hierarchy and 2) employing effective promotion and eviction methods between cache levels.}
{Our extensive experiments conducted {with} a {real} implementation of the proposed {two-level storage caching} architecture show that \techname{} {provides} 45\% higher performance{, compared to the state-of-the-art caching schemes in virtualized platforms,} while improving {both} {cache size} and SSD endurance by 51.7\% and 33.8\%, respectively.}
	
\end{abstract}}

}

\maketitle

\section{Introduction}
\label{sec:introduction}
\sloppy
Virtualized platforms are widely used in large scale data centers {to} provide significantly improved {availability} and flexibility.
In such platforms, multiple \emph{Virtual Machines} (VMs) provide different services on shared hardware.
VM management and resource partitioning between VMs are performed by {the} {hypervisor,} which plays {a major} role in the virtualized platforms and {aims to} {maximize} performance and {system} utilization.
The key advantages of virtualized platforms that {are} continuously {evolving} in computing industry are: 1) high flexibility due to {the ability to run} multiple VMs with different \emph{Operating Systems} (OS), 2) high resource utilization, 3) {resource} isolation {between different VMs}, and 4) {allocation of} dynamically adjustable resources to the VMs \cite{virtualization}.

\begin{figure}[!h]
	\centering
	\includegraphics[scale=1]{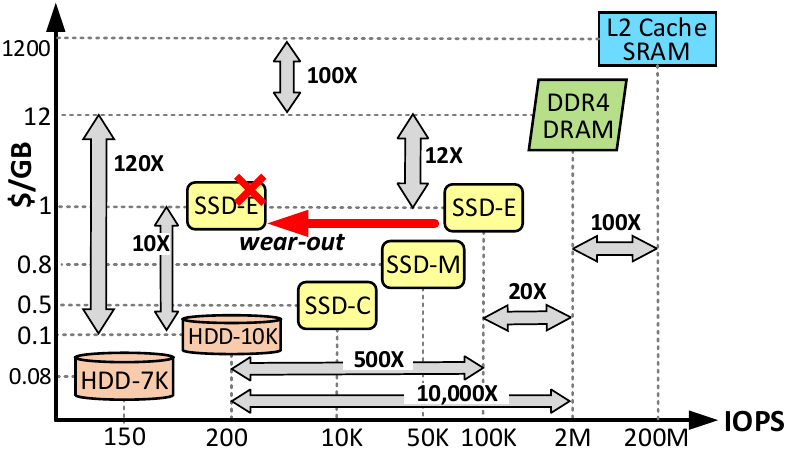}
	\caption{Comparison of different storage devices in terms of cost {(\$/GB)} and performance (IOPS). SSD-C: Consumer SSD, SSD-M: Midrange SSD, SSD-E: Enterprise SSD, DDR4 DRAM: Double Data Rate 4$^{th}$ generation, {HDD-7K: HDD with 7K \emph{Revolutions Per Minute} (RPM), HDD-10K: HDD with 10K RPM \cite{storagereview}}.}
	\label{fig:tradeoff}
\end{figure}

\sloppy
The {growing popularity} of I/O intensive applications in data {centers,} such as \emph{Online Transaction Processing} (OLTP), banking, {data analysis}, and {other} big data {workloads,} {greatly increases} the demand for {high-performance} storage subsystems.
Existing storage systems {still} mainly consist of high-capacity and low-performance \emph{Hard Disk Drives} (HDDs) with the {goal} of storing {very} large {amounts of} data {at low cost}.
Such storage systems have become the performance bottleneck in large-scale data centers due to the growing performance gap between HDDs and processing elements in enterprise servers (as depicted {by \emph{I/O Operations Per Second} (IOPS) performance} in Fig. \ref{fig:tradeoff}).

To alleviate the performance shortcomings of such HDD-based storage systems, enterprise {manufacturers,} such as Dell EMC, HP, and NetApp \cite{emc_fast_cache-online,hp_smart_cache-online,netapp_ssd_cache-online}{, and} {various research works on} emerging storage {architectures,} such as \cite{ahmadian2018eci,hystor,reca, vcacheshare, scave, centaur,janus,mercury, migratetossd,xia2016high,matthews2008intel,kim2011hybridstore,nitro,azor,xia2015flash,Yu:SSD-benifits,tica,lbica}{, propose} {\emph{I/O caching}} schemes {that} accelerate the response time of I/O requests.
High-performance storage devices such as \emph{Dynamic Random Access Memories} (DRAMs) and \emph{Solid-State Drives} (SSDs) can be employed in the {I/O} caching layer to alleviate the low performance of HDDs. 
As depicted in Fig. \ref{fig:tradeoff}, DRAM provides the highest performance among {many} types of storage/memory devices \cite{Ousterhout:dram-cache} {and thus} seems to be the best candidate {to} {employ} in {the} I/O {caching} layer. 
However, {DRAM is expensive, and} using {\emph{volatile}} DRAM as a cache comes with reliability {hurdles for write requests.}
{A DRAM-based I/O cache} requires additional battery backups to {correctly} maintain {the data buffered in DRAM} {to survive} power outages {and system failures}.
Compared to DRAM, enterprise SSDs {have} about 12X lower cost and offer 20X less performance but they provide 500X higher performance {than} HDDs.
In {contrast} to DRAM, SSDs are non-volatile and typically do not suffer from data loss due to  power outages, but they  support {\emph{only}} a limited number of reliable writes \cite{Sampson:flash-wearout,ahmadian-ssd-rel-date,cai2017errors,cai2012error,cai2017error,luo2018improving}. 
{Thus, the} main shortcoming of SSDs is their endurance limitation where both performance and reliability are significantly degraded when the number of committed writes exceeds the endurance limit (i.e., when {the} SSD wears out{, as} shown in Fig. \ref{fig:tradeoff}). In addition, {frequent} replacement of such expensive SSDs imposes significant cost and reliability issues\footnote{Due to the {degraded mode} of \emph{Redundant Array of Independent Disks} (RAID) \cite{patterson1988case} during {the} rebuild process in SSDs.} {in} storage systems.
Recent {studies,} such as \cite{ahmadian2018eci,hystor,reca, vcacheshare, scave, centaur,cloudcache,janus,mercury, migratetossd,xia2016high,matthews2008intel,kim2011hybridstore,nitro,azor,xia2015flash}{, propose} I/O caching schemes based on SSDs.
Among such studies, only few are applicable to virtualized platforms \cite{ahmadian2018eci,vcacheshare,scave,centaur,cloudcache}.
Such I/O caching schemes
aim to either {only} improve performance without considering {the} endurance of the SSD \cite{hystor,scave,centaur,cloudcache,janus,mercury,migratetossd} or attempt to improve performance with minimum {overhead on} SSD endurance \cite{ahmadian2018eci,reca,vcacheshare}. In addition, such studies suffer from two main shortcomings: 1) their performance improvement {is} limited by the {high} SSD latency {(relative to DRAM)} and 2) their endurance-{preserving} techniques lead to performance degradation.

Another approach to {improving the} performance of storage systems is to use hybrid {techniques that} take advantage {of} {\emph{both}} DRAM and {SSDs} in the I/O caching layer.
{Multiple industrial approaches (e.g., the} Dell EMC FAST cache \cite{emc_fast_cache-online} and ZFS L2ARC cache \cite{ryu2012state,zfs-gregg,zfs-gregg-blog,zfs-fs}) and academic studies \cite{ucache,molar,dulloor2016data,lee2014clock,dhiman2009pdram,ramos2011page,zhang2016unified,liu2017hardware,mittal2016survey,dong2009leveraging,mogul2009operating,gamell2013exploring,khouzani2015improving,rodriguez2013reducing,hirofuchi2016raminate,baek2014designing,liu2016memos} employ  two levels of {I/O caching} {using both} DRAM and {emerging \emph{Non-Volatile Memories} (NVMs)/SSDs} in their storage architectures.
Such caching schemes{, however,} suffer from several important shortcomings: 
{1) They {cannot provide} cache space partitioning across VMs, which results in poor performance and {high} cost in virtualized platforms.}
2) Buffered data in DRAM cache is {at} risk of {loss due to} sudden power {outage;} additional battery backups add significant cost overhead to the caching scheme. 
3) The employed {mechanisms} for detecting hot data blocks are not accurate {since they do not consider any} workload characteristics{, such} as reuse distance of the accesses.
{4) {Past works} do not provide write policy management at DRAM or SSD levels and most schemes {configure} both {cache} levels to {use the} \emph{Write Back} (WB)}\footnote{In a WB cache, {the} data block is written into the cache but the corresponding location in {the hard} disk is updated only when the data block is evicted from {the} cache, or {at} specific time intervals.} policy. 
5) Such schemes only focus on detecting and promoting {\emph{hot}} data blocks to the cache while there is no specified method to detect and evict {\emph{cold} data blocks} (i.e., {least recently accessed data blocks})  from cache.

Most recent cache space partitioning schemes in virtualized platforms employ reuse distance analysis {to estimate each VM's} cache size{, but} they neglect key parameters such as \emph{request type} and \emph{cache write policy} \cite{centaur,ahmadian2018eci}. The state-of-the-art scheme, ECI-Cache \cite{ahmadian2018eci}, proposes \emph{Useful Reuse Distance} (URD) and considers the \emph{request type} in reuse distance calculation.
{However, it} neglects the impact of \emph{cache write policy} {on the reuse distance calculation, and, as a result, it} {over-estimates} {the} cache sizes for caches with write policies other than WB and \emph{Write Through} (WT).
ECI-Cache estimates {the} cache {sizes} of the VMs based on the URD metric and assigns {a cache write policy on a per-VM basis}. {As such, it results} in improved SSD‌ endurance and performance-per-cost.
{The} URD {metric} is optimized for caches with WB and WT policies{. It} {\emph{over-estimates}} the cache size {for} caches with other write policies (e.g., \emph{Read Only} (RO)).
{If we use the} URD metric in {a cache} with {the} RO policy, cache blocks would be {allocated} for \emph{write} requests {that} would not be buffered in the RO cache{, and thus the allocated blocks remain \emph{unused}}.

In this paper, we first propose a new metric, called \emph{\URDPfull{}} (\URDP{}), which considers the impact of {the} \emph{cache write policy} in addition to \emph{request type} in reuse distance calculation. \URDP{}{, unlike URD,} does {\emph{not}} reserve cache blocks for accesses {that} would not be supplied by the cache {and,} hence, {allocates} much smaller cache space {for a VM,} compared to URD, while preserving {the VM's} performance.
Second, we propose the \emph{\shfullname{}} (\techname{}) for virtualized platforms.
\techname{} employs {\emph{two levels}} of cache{, with} DRAM at the first and SSD‌ at the second level.
Using the \URDP{} metric, the proposed architecture  effectively partitions the space of both levels of cache between VMs and improves both performance and endurance.


{In our proposal, we first  assign an {effective} and fixed write policy to each cache level {that} {provides high} reliability and {high} endurance in DRAM and SSD levels, respectively.} Second, using {the} \URDP{} {metric,} we partition the cache space by assigning {an} efficient cache size to each VM{, which  maximizes} performance-per-cost. Third, we propose promotion and eviction methods to effectively transfer \emph{popular} (i.e., \emph{hot}) and \emph{unpopular} (i.e., \emph{cold}) data blocks between {the two cache} levels.
Unlike previous I/O caching schemes, our proposed two-level cache does {\emph{not}} evict popular data blocks from the cache until they become unpopular.
{Via} online monitoring of I/O characteristics of the running workload, we extract several metrics such as 1) frequency, 2) type, and 3) \URDP{} of accesses in the running VMs. The extracted information {is} used to 1) estimate {an} efficient cache size for {each VM} and 2) detect \emph{popular} and \emph{unpopular} data blocks.
To summarize, our proposed two-level I/O {caching} scheme aims to: 1) enhance both read and write performance, 2) overcome the reliability issues of {the} volatile DRAM cache, 3) improve {the lifetime} of the SSD cache, and 4) reduce {the} {overall} cost of the {two-level} cache (in terms of cache size {allocated to each VM}).


{We implement our proposed two-level I/O {caching mechanism} on a {real virtualized} system including: 1) per-VM cache size estimation, 2) write policy assignment, 3) popular/unpopular data block detection, and 4) promotion/eviction {mechanisms}.} 
{We {evaluate} \techname{} with more than 10 VMs running real application workloads from SNIA \cite{msr,narayanan2008write}. {Our} experimental results  show that the proposed architecture {provides} 45\% lower I/O latency while {also} improving effective {cache size and} SSD endurance by 51.7\% and 33.8\%, respectively, compared to {the best} state-of-the-art caching {scheme} in virtualized platforms.}

{We} make the following contributions.
\begin{itemize}[leftmargin=*]
	
	\item We propose a novel {\emph{two-level I/O cache}} for virtualized platforms, {using} DRAM and SSD{, that} resolves reliability, endurance, and performance issues {\emph{without}} any additional high-cost peripherals such as battery backups or internal disks.
	
	\item We propose a new metric, \emph{\URDPfull{}} (\URDP{}), which refines the concept of {useful} reuse distance calculation based on \emph{cache write policy}. {This} metric does {\emph{not}} allocate cache space for accesses {that} would not be served by {the} cache{. Thus, it} assigns {a} smaller cache space to {each} {VM,} resulting in {lower} cost.

	\item We {propose a mechanism to assign} efficient write policies for different cache levels to balance endurance and reliability of the I/O cache. {Our mechanism enables} read {requests to be} supplied from DRAM and write requests {to be} supplied from {the} SSD. 
	
	\item {Via} online characterization of I/O requests, {our technique} effectively {determines} \emph{popular} and \emph{unpopular} data blocks. Our proposed promotion and eviction methods buffer and hold data blocks in the cache {while} they {remain} \emph{popular}.
	
	\item We implement the proposed two-level I/O cache in a real system and evaluate the performance and endurance of our scheme by performing {extensive} experiments with more than 10 VMs running real application workloads on an open-source hypervisor, QEMU \cite{qemu}. 
	
	\item {{We find \techname{} to provide higher performance, higher efficiency, and higher endurance than the state-of-the art I/O caching schemes for virtualized platforms \cite{ahmadian2018eci}.}}
	
\end{itemize}


\sloppy

\vspace{-0.7em}
\section{Related Work}
\label{sec:related}
In this section, we first {describe and analyze} the previous I/O caching architectures in virtualized platforms. Second, we analyze the existing I/O cache architectures in two {groups:} 1) single level and 2) multi {level,} which are mainly employed in {\emph{non-virtualized}} platforms.

\subsection{I/O Caching in Virtualized Platforms}
S-CAVE \cite{scave}, vCacheShare \cite{vcacheshare}, Centaur \cite{centaur}, and ECI-Cache \cite{ahmadian2018eci} are the most state-of-the-art hypervisor-based I/O caching schemes in virtualized platforms. Such schemes employ a {\emph{single level}} of SSD cache and mainly focus on dynamic and efficient cache space partitioning between running VMs.
S-CAVE \cite{scave} {uses} the \emph{Working Set Size} (WSS) of the VMs {for} cache {space} estimation. To preserve the reliability {of write requests}, the write policy of the SSD cache is set to \emph{Write Through} (WT). Such cache size estimation (based on WSS) is deprecated and fails {to accurately} {estimate} the cache {space needed} for workloads with sequential access {patterns} \cite{ahmadian2018eci}.

vCacheShare \cite{vcacheshare} estimates the cache size of the VMs based on locality and reuse intensity (i.e., workloads' burstiness). vCacheShare considers both reliability and endurance by applying \emph{Write Around} (WA) or \emph{Read Only} (RO) write {policies} which {direct} the write requests to the disk subsystem and {\emph{only}} {improve} the performance of read accesses {via caching}. The cache size estimation scheme used in vCacheShare is  applicable to CPU {caches,} {but its} assumptions ({e.g.,} reuse intensity) {\emph{cannot}} be applied in I/O caches \cite{centaur}. vCacheShare does {\emph{not}} improve the performance of write requests.

Centaur \cite{centaur} assigns cache {space to} the running VMs based on reuse distance analysis{, which is} an effective approach {to} cache size estimation. The write policy of the SSD cache is simply set to \emph{Write Back} (WB).
Such {a} cache size estimation scheme does not consider {either the} \emph{request type} {or the} \emph{cache write policy}{, which} leads to {over-estimation of} the cache sizes{, negatively impacting cost}. In addition, by using {the} WB policy, Centaur {negatively affects} both {storage} reliability and {the} endurance of the SSD cache.

ECI-Cache \cite{ahmadian2018eci} is the latest state-of-the-art I/O caching scheme for virtualized {platforms,} which {overcomes} the shortcomings of previous schemes. {It} proposes {the} \emph{Useful Reuse Distance} {(URD) {metric},} which considers \emph{request type} in cache size estimation and reduces the allocated cache space {while} preserving performance{, enabling lower cost}. ECI-Cache dynamically assigns {either the} WB {or the} RO policies {for each VM} and {thus} provides {higher} performance and endurance {for} the SSD cache.
{There are four issues with ECI-Cache, which we aim to {overcome with our new design}.}
{\textbf{First}, }
{ECI-Cache dynamically assigns RO and WB policies on the {VM's} cache, while} URD-based cache size estimation does {\emph{not}} consider {the} \emph{cache write policy} and only works {for} WB {and WT} caches.
URD reserves cache blocks for the accesses {that} would {\emph{not}} be served by {the} cache due to different cache write policies, and therefore, it {\emph{over-estimates}} the size of the caches {using} other {policies,} such as RO. For instance, in {an} RO cache, URD reserves cache blocks for write accesses {that} would {\emph{not}} be buffered in the cache.
{\textbf{Second}, }ECI-Cache is {only} {a \emph{one-level} {(i.e., SSD-only)}} I/O cache where the performance improvement is limited {by} the performance of the SSD.
{\textbf{Third}, }ECI-Cache updates the cache content on each cache miss and {therefore} imposes {a} large number of unnecessary writes into the SSD cache.
{\textbf{{Fourth}}, }{ECI-Cache's cache update} scheme {is supposed to promote any missed data blocks into the cache. Such scheme} may evict a \emph{hot} data block to promote a data block without any future references into the {cache,} resulting in performance degradation.

To summarize, existing hypervisor-based I/O caching architectures suffer from three {major} shortcomings: 1) they  employ {only} {an} SSD in the {caching} layer{, and thus their} performance improvement is limited {by} the performance of {the} SSD {(which is much lower than that of DRAM)}, 2) they fail in cache size estimation {under} different workloads and different cache write policies, and 3) they fail in balancing performance and endurance.
\subsection{I/O Caching in Non-Virtualized Platforms}
\subsubsection{Single Level I/O Caches}
{A relatively} high-performance {memory device,} such as DRAM or SSD is employed in {the} {caching} layer to close the performance gap between {the} processing {units} and {the} storage subsystem (which is mainly composed of {low-performance} HDDs) \cite{argon,migratetossd,nitro,hystor,reca,azor,janus}. Such I/O caching schemes improve {the} performance of I/O requests by providing efficient cache space allocation and configuration. {However}, they neglect other key parameters such as {storage} reliability and endurance of the SSDs.
Argon \cite{argon} aims to efficiently partition the DRAM cache space between {different} services to maximize the cache hit ratio.
Janus \cite{janus} allocates {SSD} space based on the ratio of  read accesses of the workload and aims to improve the performance of read requests.
Hystor \cite{hystor} employs {an} SSD cache to improve {the} performance of read and write accesses and aims to identify and buffer the data blocks{, which helps} {in improving} the hit ratio.
ReCA \cite{reca} characterizes the workloads of running applications and provides a per-application cache configuration. This scheme aims to improve both performance and endurance by assigning different cache policies {to different applications, but it cannot minimize} the number of  SSD writes (mainly the writes due to read misses) {because it uses} a single SSD in the {caching} layer.
{ReCA is able to reconfigure 1) cache line size, 2) write policy, and 3) eviction policy. This scheme allocates cache space globally and neglects the cache management in case of running multiple workloads (i.e., services). Such scheme cannot be applied in virtualized platforms, where multiple VMs are running on the system. Since this approach only employs SSDs in the caching layer, the performance improvement is limited by the highest performance of the SSDs.}
SHARDS \cite{shards} presents a hashed approximate reuse distance sampling scheme to be used in cache size estimation. This scheme improves the performance of reuse distance calculation without considering requests type and cache write policy in reuse distance calculation. 

\subsubsection{Multi Level I/O Caches}
Such schemes employ both SSD and DRAM in the {I/O caching} {layer}.\footnote{Many recent works examine DRAM cache design \cite{qureshi2009scalable,meza2012enabling,li2017utility,agarwal2015page,agarwal2017thermostat,goglin2016exposing,salkhordeh2016operating} for hybrid main memory systems, e.g., those combining DRAM and \emph{Phase Change Memory} (PCM) \cite{lee2009architecting, lee2010phase, qureshi2009scalable} or  \emph{Spin-Transfer Torque Magnetic Random-Access Memory} (STT-MRAM) \cite{kultursay2013evaluating}. Our work is significantly {different, since} none of these recent works 1) are applicable to I/O caching in virtualized storage systems (as they focus on main memory), 2) develop the new reuse distance metric we introduce and evaluate. As such, our work is orthogonal to such hybrid main memory system designs, and the reuse distance metric we introduce can be useful within the hybrid main memory system context -- an avenue we leave for future work.}
uCache \cite{ucache} provides a simple two-level I/O cache design by employing DRAM and SSD. 
In this scheme, data blocks are kept in DRAM {until they are evicted, and after eviction they} are moved to the SSD.
uCache improves the process of demoting data blocks from DRAM to SSD by aggregating {a} large number of small writes {and,} hence, improves the endurance of the SSD. {However}, {write-pending} data blocks are kept in {the} {\emph{volatile}} DRAM {cache,} which results in data loss in case of power failure. 
Molar \cite{molar} presents a {simulator-based} hybrid I/O cache without any improvement in eviction/promotion methods. This scheme consists of DRAM, SSD, and HDD tiers and migrates data blocks between them. Molar aims to predict the future accesses and decides which data block should be evicted from DRAM {based on this prediction}. {This} scheme does {\emph{not}} employ any cache write policy management on different tiers and hence cannot preserve 1) the reliability of buffered data blocks in {the} volatile DRAM {cache,} and 2) {the} endurance of the SSD cache.

{In \cite{tica}, DRAM, \emph{Write Optimized SSD} (WO-SSD), and \emph{Read Optimized SSD} (RO-SSD) are used in the caching layer. Write requests are buffered in both DRAM and WO-SSD, read misses are promoted to DRAM, and evictions from DRAM are directed into the RO-SSD. The main shortcomings of this architecture are:  1) Write requests are directed to both DRAM and WO-SSD, hence, they experience the write latency of the SSD. Using DRAM to buffer the write accesses has \emph{no} impact on the performance of writes and only helps to improve the latency of future reads that may hit in DRAM (i.e., RAW accesses). The limited DRAM space and small ratio of RAW accesses lead to very small performance improvement by this architecture. 2) This architecture prevents buffering write accesses in the RO-SSD (to preserve the lifetime of RO-SSD), but on the other hand, evictions from the DRAM and promotions of read misses are performed on RO-SSD which impose extra writes and have a negative impact on the limited lifetime of RO-SSDs. 3) This method is a \emph{global} cache which cannot assign efficient cache space for the running services, and hence, is not applicable in virtualized platforms.}

\begin{figure}[!htb]
	\centering
	\subfloat[L2ARC]{\includegraphics[width=.16\textwidth]{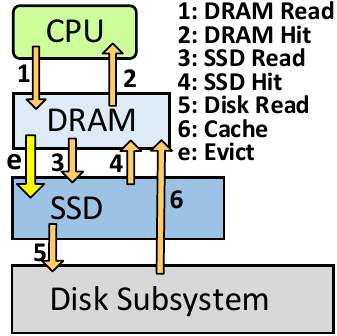}%
		\label{fig:l2arc}}
	\hfil
	\hspace{-0.4em}
	\subfloat[FAST (Read)]{\includegraphics[width=.16\textwidth]{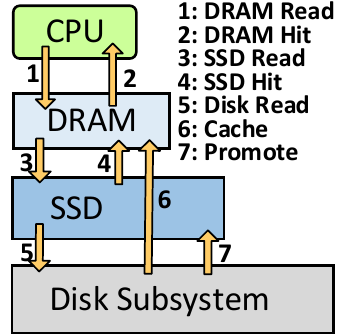}%
		\label{fig:fast-read}}
	\hfil
	\hspace{-0.4em}
	\subfloat[FAST (Write)]{\includegraphics[width=.16\textwidth]{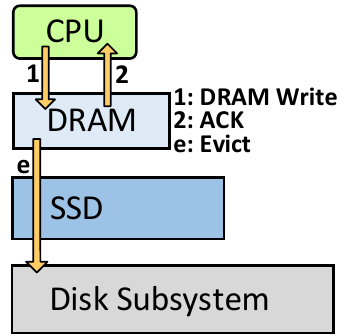}%
		\label{fig:fast-write}}
	
	\caption{Architecture of two-level I/O caches {employed in enterprise storage systems}.}
	\label{fig:prev_2level}
\end{figure}

\emph{Zettabyte File System} (ZFS) \emph{Level 2 Adaptive Replacement Cache} (L2ARC) \cite{ryu2012state,zfs-gregg,zfs-gregg-blog,zfs-fs} is a file system level {cache,} which improves {upon} \emph{Adaptive Replacement Cache} (ARC) \cite{megiddo2003arc} by employing {an} SSD between DRAM and HDD (i.e., disk subsystem), thereby reducing the latency of read misses. L2ARC works in a simple \emph{First In First Out} (FIFO) mode and only improves the performance of read requests.
As shown in Fig. \ref{fig:l2arc}\footnote{L2ARC does {\emph{not}} buffer write requests {in {the} SSD} and, hence, we do not provide the flow of write accesses in Fig. \ref{fig:prev_2level}.}, read requests may be supplied from DRAM or SSD (i.e., hit{),} {and} read misses are supplied from {the} disk subsystem and promoted {into} DRAM. L2ARC predicts to-be-evicted data blocks from DRAM and pushes them to the SSD (before they {have to be} evicted).
{The major shortcomings of L2ARC are: 1)} {it} lacks an efficient promotion/eviction {method:} only a simple FIFO manages the contents of the SSD,
2) {it} {could waste}  the SSD space by pushing data blocks from DRAM to SSD before their eviction,
3) by keeping evicted blocks in the SSD, L2ARC mainly improves the performance of {\emph{Read After Read} (RAR)} accesses with a reuse distance larger than {the} DRAM size (i.e., non-popular blocks{),} which has {a} {relatively small} impact on overall performance.

Dell EMC \emph{Fully Automated Storage Tiering} (FAST) cache \cite{emc_fast_cache-online} {is} an enterprise approach {that} employs  a two-level I/O cache in storage products.
{This} scheme {sets} the write policy of {the} DRAM cache (called SP\footnote{SP: Storage Processor} cache) to \emph{Write Back} (WB) to accelerate the performance of write requests {and uses} battery backups to maintain {reliability in the presence of power and system failures}.
{The} write policy of the SSD‌ cache is {also} set to WB.
Fig. \ref{fig:fast-read} and Fig. \ref{fig:fast-write} show how FAST handles read and write requests.
{It} employs {a very simplistic} method to identify hot data blocks where a block with more than three accesses in {a} recent time interval is identified as hot.
{Hot blocks are} promoted to the SSD {cache.} {There} is no specified rule for evicting data blocks from cache.
{Thus, FAST may evict} hot data blocks early from {the} cache (due to promoting new data blocks{),} which affects both performance and endurance of the SSD. 
In this case, {promotion of the evicted hot data block again} imposes additional write operations on the SSD.

Table \ref{table:prv_table} summarizes the previously discussed architectures.
{Among all} previous {proposals}, uCache \cite{ucache}, L2ARC \cite{ryu2012state,zfs-gregg,zfs-gregg-blog,zfs-fs}, and FAST \cite{emc_fast_cache-online} as two-level {caching} schemes and ECI-Cache \cite{ahmadian2018eci} as a hypervisor-based I/O caching scheme are close to our proposal. 
{However, the previous two-level caching schemes neglect cache write policy assignment on different cache levels and hence, fail {to} {{provide} storage} reliability and {high} endurance {for} the SSD cache.
L2ARC works in the file system level and only improves read performance without considering the endurance of the SSDs. L2ARC fails in detecting popular blocks and does not provide efficient eviction/promotion methods where the cache is managed in a simple FIFO mode.
FAST cache keeps write pending data blocks in volatile DRAM cache and employs additional battery backups to provide reliability and hence imposes additional cost. However, such scheme cannot guarantee the reliability of write pending data blocks. In addition, FAST cache employs a simple method to recognize hot data blocks which fails in different types of workloads. 
uCache keeps both read and write data blocks in DRAM. Thus, a power failure leads to data loss. On the other hand, the DRAM cache becomes polluted due to caching all requests in such small space. In addition, uCache does not consider the popularity of data blocks in the proposed caching scheme.}
ECI-Cache {dynamically assigns different write policies on {different} VMs {caches} but} does not take into account the impact of cache write policy on cache size estimation and hence, overestimates cache sizes for the VMs {that use} write policies other than WB. ECI-Cache imposes additional write operations to the SSD cache by updating the cache on each miss operation. In addition, {{it} is one level: it employs only an SSD, no DRAM.} {Buffering} both read and write requests {in the SSD} reduces the SSD endurance.
\begin{table}[!h]
\centering
\caption{Summary of previous architectures (WSS: Working Set Size, RI: Reuse Intensity, TRD: Traditional Reuse Distance, URD: Useful Reuse Distance, and \URDP: \URDPfull).}
\label{table:prv_table}
\scriptsize
\begin{tabular}{|l||c|c|c|c|c|c|c|}
\hline
\centering Architecture & 
\begin{turn}{90}Industrial\end{turn} & \begin{turn}{90}Virtualized\end{turn} & \begin{turn}{90}\begin{tabular}[c]{@{}l@{}}Cache Space\\ Partitioning\end{tabular}\end{turn} & \begin{turn}{90}Multi Level\end{turn}    & 
\begin{turn}{90}\begin{tabular}[c]{@{}l@{}}Write Policy\\ Management\end{tabular}\end{turn} & \begin{turn}{90}\begin{tabular}[c]{@{}l@{}}Endurance\\Improvement\end{tabular}\end{turn} & 
\begin{turn}{90}\begin{tabular}[c]{@{}l@{}}Reliability\\Improvement\end{tabular}\end{turn}\\
 \hline\hline
S-CAVE \cite{scave}                                                    & \XSolidBold         & \CheckmarkBold         & WSS                                                       & \multicolumn{1}{c|}{\begin{tabular}[c]{@{}c@{}}\XSolidBold\\ (SSD)\end{tabular}}       & \XSolidBold                                                                & \XSolidBold                                                        & \CheckmarkBold                                                         \\ \hline
vCS \cite{vcacheshare}                                              & \XSolidBold         & \CheckmarkBold         & RI                                                        &\multicolumn{1}{c|}{\begin{tabular}[c]{@{}c@{}}\XSolidBold\\ (SSD)\end{tabular}}       & \XSolidBold                                                                & \CheckmarkBold                                                       & \CheckmarkBold                                                         \\ \hline
Centaur \cite{centaur}                                                   & \XSolidBold         & \CheckmarkBold         & TRD                                                       & \multicolumn{1}{c|}{\begin{tabular}[c]{@{}c@{}}\XSolidBold\\ (SSD)\end{tabular}}       & \XSolidBold                                                                & \XSolidBold                                                        & \XSolidBold                                                          \\ \hline
ECI-Cache \cite{ahmadian2018eci}                                                & \XSolidBold         & \CheckmarkBold         & URD                                                       & \multicolumn{1}{c|}{\begin{tabular}[c]{@{}c@{}}\XSolidBold\\ (SSD)\end{tabular}}       & \CheckmarkBold                                                               & \CheckmarkBold                                                       & \XSolidBold                                                          \\ \hline
Argon \cite{argon}                                                    & \XSolidBold         & \XSolidBold          & {\begin{tabular}[c]{@{}c@{}}Acc.\\ Based\end{tabular}}                                                & \multicolumn{1}{c|}{\begin{tabular}[c]{@{}c@{}}\XSolidBold\\ (DRAM)\end{tabular}}      & \XSolidBold                                                                & -                                                         & \XSolidBold                                                          \\ \hline
Janus  \cite{janus}                                                   & \XSolidBold         & \XSolidBold          & {\begin{tabular}[c]{@{}c@{}}Read \\ Acc.\\ Based\end{tabular}} & \multicolumn{1}{c|}{\begin{tabular}[c]{@{}c@{}}\XSolidBold\\ (SSD)\end{tabular}}       & \XSolidBold                                                                & \XSolidBold                                                        & \XSolidBold                                                          \\ \hline
Hystor  \cite{hystor}                                                 & \XSolidBold         & \XSolidBold          & {Global}                                                        & \multicolumn{1}{c|}{\begin{tabular}[c]{@{}c@{}}\XSolidBold\\ (SSD)\end{tabular}}       & \XSolidBold                                                                & \XSolidBold                                                        & \XSolidBold                                                          \\ \hline
ReCA \cite{reca}                                                   & \XSolidBold         & \XSolidBold          & {Global}                                                        & \multicolumn{1}{c|}{\begin{tabular}[c]{@{}c@{}}\XSolidBold\\ (SSD)\end{tabular}}       & \CheckmarkBold                                                               & \CheckmarkBold                                                       & \XSolidBold                                                          \\ \hline
 \cite{tica}                                                   & \XSolidBold         & \XSolidBold          & {Global}                                                        & \multicolumn{1}{c|}{\begin{tabular}[c]{@{}c@{}}\CheckmarkBold\\ (DRAM+SSD)\end{tabular}}       & \XSolidBold                                                               & \CheckmarkBold                                                       & \XSolidBold                                                          \\ \hline
uCache  \cite{ucache}                                                  & \XSolidBold         & \XSolidBold          & {Global}                                                        & \multicolumn{1}{c|}{\begin{tabular}[c]{@{}c@{}}\CheckmarkBold\\ (DRAM+SSD)\end{tabular}} & \XSolidBold                                                                & \XSolidBold                                                        & \XSolidBold                                                          \\ \hline
\begin{tabular}[c]{@{}l@{}}L2ARC\\\cite{ryu2012state,zfs-gregg}\\\cite{zfs-gregg-blog,zfs-fs}\end{tabular}                                                    & \CheckmarkBold        & \XSolidBold          & {Global}                                                        & \multicolumn{1}{c|}{\begin{tabular}[c]{@{}c@{}}\CheckmarkBold\\ (DRAM+SSD)\end{tabular}} & \XSolidBold                                                                & \XSolidBold                                                        & \XSolidBold                                                          \\ \hline
FAST  \cite{emc_fast_cache-online}                                                    & \CheckmarkBold        & \XSolidBold          & {Global}                                                        & \multicolumn{1}{c|}{\begin{tabular}[c]{@{}c@{}}\CheckmarkBold\\ (DRAM+SSD)\end{tabular}} & \XSolidBold                                                                & \CheckmarkBold                                                       & \XSolidBold                                                          \\ \hline
\begin{tabular}[c]{@{}l@{}}\textbf{\techname{}}\\ (Proposed)\end{tabular} & \XSolidBold         & \CheckmarkBold         & \URDP                                                       & \multicolumn{1}{c|}{\begin{tabular}[c]{@{}c@{}}\CheckmarkBold\\ (DRAM+SSD)\end{tabular}} & \CheckmarkBold                                                               & \CheckmarkBold                                                       & \CheckmarkBold                                                         \\ \hline
\end{tabular}
  \vspace{-0.3em}
\end{table}

\section{Motivation and Illustrative Example}
\label{sec:motivation}
I/O caches are employed to enhance the performance of storage systems.
{High-locality} data blocks are kept in {the} cache to serve future requests {to them faster}  and hence reduce the response time significantly.
Currently, enterprise storage systems employ {high-performance} SSDs in the I/O {caching} layer. {SSDs provide more than 500X {the} performance {provided by HDDs} {(in terms of IOPS for random requests)} but {SSDs} wear out fast due to write operations.} {Depending} on the cache \emph{write policy}, both read and write requests from the application layer may impose write operations on the SSD.
Four different write policies can be {used on} the cache. We elaborate on how 1) requests are supplied by {the} cache, and 2) performance, reliability, and endurance are affected by different write policies:
\begin{itemize}[leftmargin=*]
	
	\item \textbf{\emph{Write Back} (WB)} (with write allocate and no read through\footnote{{A read through cache buffers read misses in the cache, and hence, serves further read accesses to that address.}}) buffers write requests in the {I/O} cache and {writes} them back to the storage subsystem when the dirty blocks are evicted. 
	Read requests are also buffered in {a} WB cache. {Each} miss from the cache imposes an additional write operation {into} the cache {to store the read block}.
	{The WB} policy reduces write accesses to the disk subsystem and {thus} improves both read and write performance. {WB policy may lose the write-pending data before they can be written back to the} storage subsystem, on power failure. In addition, an SSD WB cache wears out fast due to {the} large number of write operations {induced on the cache}.
	
	\item \textbf{\emph{Write Through} (WT)} (with write allocate and no read through) buffers write requests {in the I/O cache} and {at the same time} also commits them to the storage subsystem. {In addition} to writes, WT cache buffers {also} read requests. {WT} does {\emph{not}} improve write performance and only {helps} read performance. WT policy preserves the reliability {of writes} since it does {\emph{not}} buffer any {write-pending} data (i.e., dirty blocks). WT policy has the same {negative} impact on {the} endurance of the SSD as {the} WB policy.
	
	\item \textbf{\emph{Write Only} (WO)} (with write allocate and read through) {handles the} write requests {in a similar way} to the WB policy. Read {requests} are {\emph{not}} buffered in the WO cache{. Hence}, this policy improves {the} performance of {only} write and \emph{Read After Write} (RAW) operations. WO {does not {incur additional}} write {operations} due to read misses {in} the cache and, therefore, significantly improves {the} endurance of {an} SSD cache.
	
	\item \textbf{\emph{Read Only} (RO)} (with no read through) {buffers} {\emph{only}} read requests and {directs} writes to the storage subsystem. {RO improves only} read performance and preserves the reliability of write requests. RO has {a} positive impact on SSD cache endurance {because it does not expose the I/O write requests to the cache}.
	
\end{itemize}

We conduct comprehensive experiments to {examine} the impact of { these} write policies on {the} performance and endurance of the SSD‌ cache  (in terms of number of write operations on the SSD).
{The results of these experiments motivate us in selecting the optimized cache policies to provide higher performance, endurance, and reliability in our proposed I/O caching architecture.}
In these experiments, we {study the use of an SSD}  as a {caching} layer for the HDD-based {storage} subsystem {with} three types of write policies: 1) WB, 2) RO, and {3) WBWO} (i.e., WB and WO).
{WT cache compared to WB (as described earlier) provides similar performance for read accesses and degraded performance for write accesses (due to simultaneous writes in both SSD‌ and HDD devices). Hence, we do not report WT cache performance in our evaluations.}

To do so, we perform experiments on a real system. The experimental setup and the running workloads are reported in Table \ref{motivation_setup}. We use an open-source EnhanceIO cache module \cite{enhanceio} to implement {our} I/O caching scheme.
{The system} is warmed up for 30 minutes and then we run the workload for one hour.

\begin{table}[!h]
	\centering
	\caption{Setup of the motivational experiments.}
	\label{motivation_setup}
	\tiny
	\begin{tabular}{|l|c|c|c|c|c|c|}
		\hline
		\multicolumn{7}{|l|}{\textbf{Hardware}}                                                                                                                                                                                                                                                                                                                                                                                      \\ \hline
		HDD                                                                   & \multicolumn{6}{l|}{\begin{tabular}[c]{@{}l@{}}6x 4TB SAS 7.2K Seagate HDD in R5 (5+1) configuration\\ Disk Partition size: 200GB\end{tabular}}                                                                                                                                                                                                  \\ \hline
		SSD                                                                   & \multicolumn{6}{l|}{\begin{tabular}[c]{@{}l@{}}4x 2TB Samsung 863a SSD in R10 (2+2) configuration\\ Disk Partition size: 50GB\end{tabular}}                                                                                                                                                                                                      \\ \hline
		DRAM                                                                  & \multicolumn{6}{l|}{128GB Samsung DDR4}                                                                                                                                                                                                                                                                                                      \\ \hline
		CPU                                                                   & \multicolumn{6}{l|}{32xIntel(R) Xeon (R) $2.1$GHz CPU core}                                                                                                                                                                                                                                                                                 \\ \hline\hline
		\multicolumn{7}{|l|}{\textbf{Software}}                                                                                                                                                                                                                                                                                                                                                                                      \\ \hline
		OS                                                                    & \multicolumn{6}{l|}{CentOS 7 (kernel version: 3.10.327)}                                                                                                                                                                                                                                                                                     \\ \hline\hline
		\multicolumn{7}{|l|}{\textbf{Workloads}}                                                                                                                                                                                                                                                                                                                                                                                     \\ \hline\hline
		\multirow{2}{*}{\begin{tabular}[c]{@{}l@{}}FIO \cite{fio_online}\\ RandRW\end{tabular}} & \begin{tabular}[c]{@{}c@{}}Req.\\ Size\end{tabular}                  & \begin{tabular}[c]{@{}c@{}}Req.\\ Type\end{tabular}               & \begin{tabular}[c]{@{}c@{}}Access\\ Pattern\end{tabular}                    & \begin{tabular}[c]{@{}c@{}}IO\\ Depth\end{tabular} & Threads & \begin{tabular}[c]{@{}c@{}}IO\\ Engine\end{tabular} \\ \cline{2-7} 
		& 8KB                                                                  & \begin{tabular}[c]{@{}c@{}}Read/Write\\ (Read: 70\%)\end{tabular} & \begin{tabular}[c]{@{}c@{}}Random\\ (distribution:\\ zipf:1.1)\end{tabular} & 16                                                 & 16      & Libaio                                              \\ \hline\hline\hline
		\begin{tabular}[c]{@{}l@{}}FileBench\\ \cite{filebench}\end{tabular} & \begin{tabular}[c]{@{}c@{}}Req.\\ Size\end{tabular}                  & \begin{tabular}[c]{@{}c@{}}Req.\\ Type\end{tabular}               & \begin{tabular}[c]{@{}c@{}}Access\\ Pattern\end{tabular}                    & \multicolumn{2}{c|}{Threads}                                 & WSS                                                 \\ \hline
		\begin{tabular}[c]{@{}l@{}}Web\\ Server\end{tabular}                  & 64KB                                                                  & Read/Write                                                        & Random                                                                      & \multicolumn{2}{c|}{100}                                     & 65GB                                                \\ \hline
		\begin{tabular}[c]{@{}l@{}}Video\\ Server\end{tabular}                & 256KB & Read                                                        & Sequential                                                                  & \multicolumn{2}{c|}{48}                                      & 120GB                                               \\ \hline
		Varmail                                                               & 64KB                                                                  & Read/Write                                                        & Random                                                                      & \multicolumn{2}{c|}{16}                                      & 62GB                                                \\ \hline
	\end{tabular}
\end{table}

\begin{figure*}[!t]
	\centering
	\subfloat[FIO-RandRW]{\includegraphics[width=.27\textwidth]{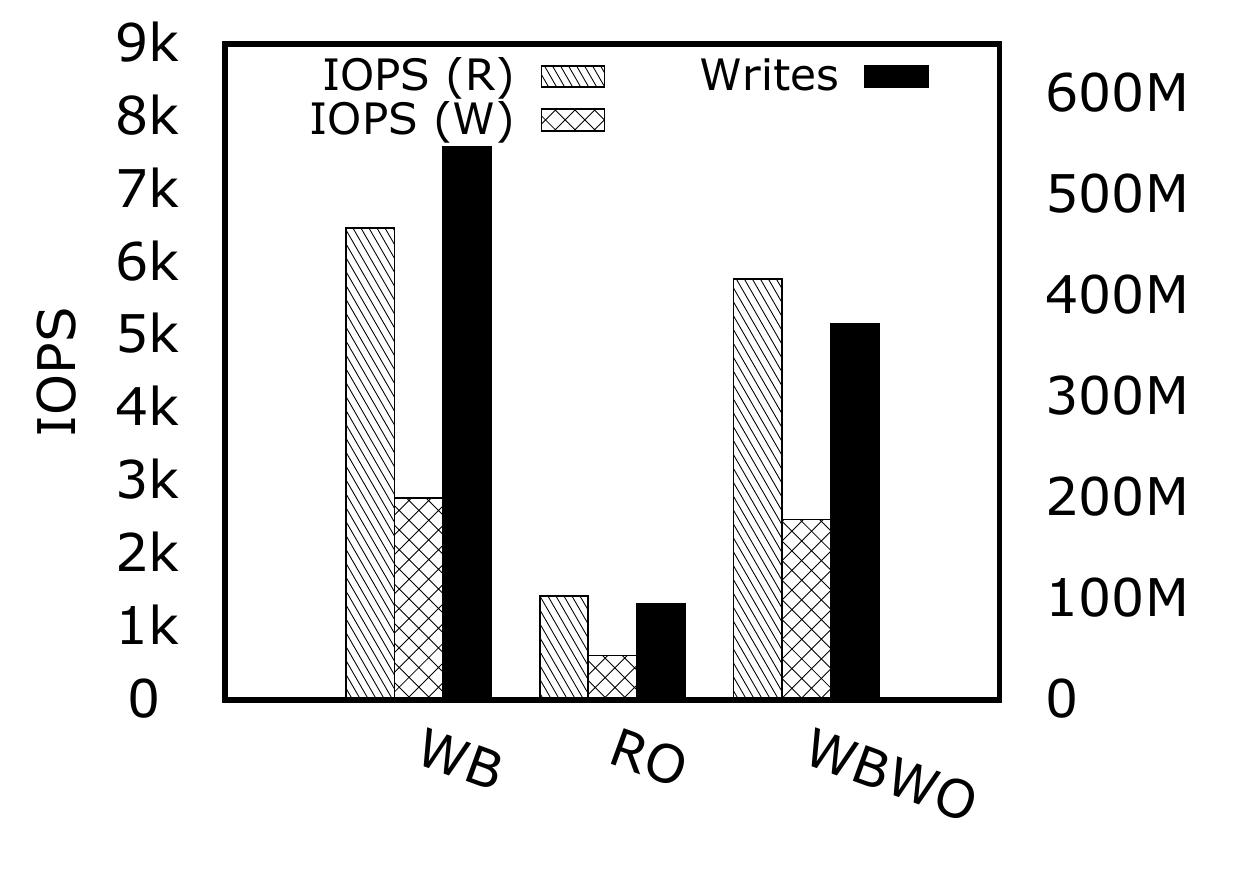}%
		\label{fig:fio-randrw}}
	\hspace{-6mm}
	\hfil
	\subfloat[Web Server]{\includegraphics[width=.27\textwidth]{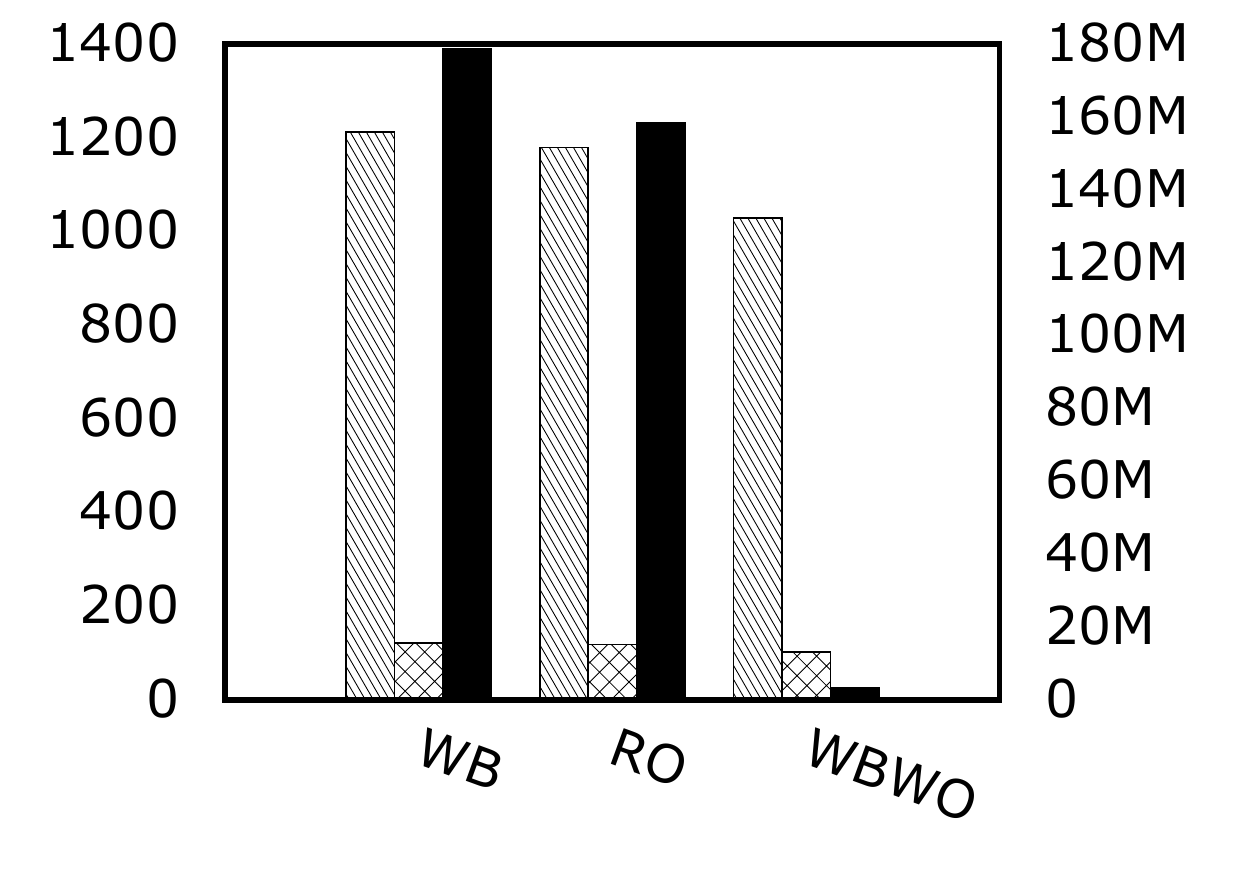}%
		\label{fig:webserver}}
	\hspace{-6mm}
	\hfil
	\subfloat[Video Server]{\includegraphics[width=.27\textwidth]{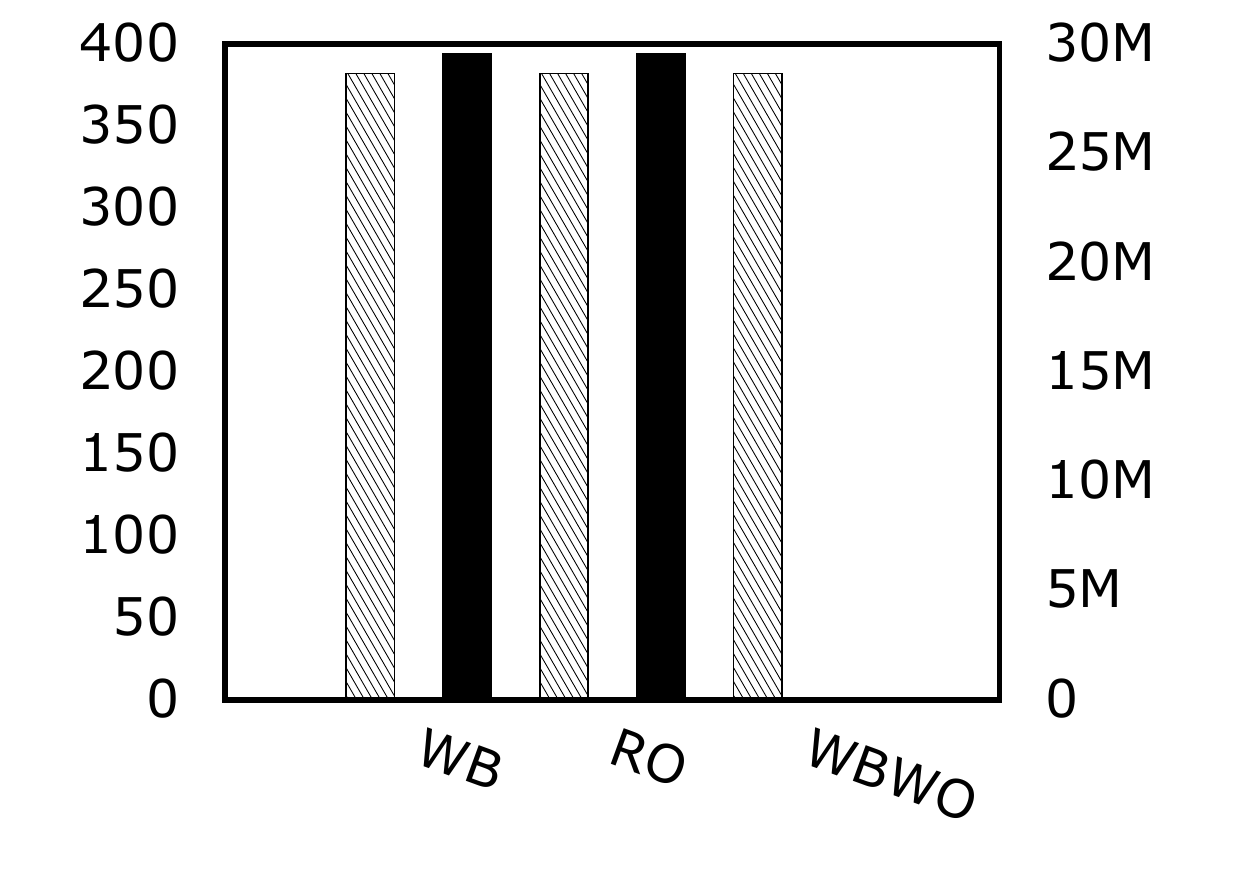}%
		\label{fig:videoserver}}
\hspace{-7mm}
\hfil
	\subfloat[Varmail]{\includegraphics[width=.27\textwidth]{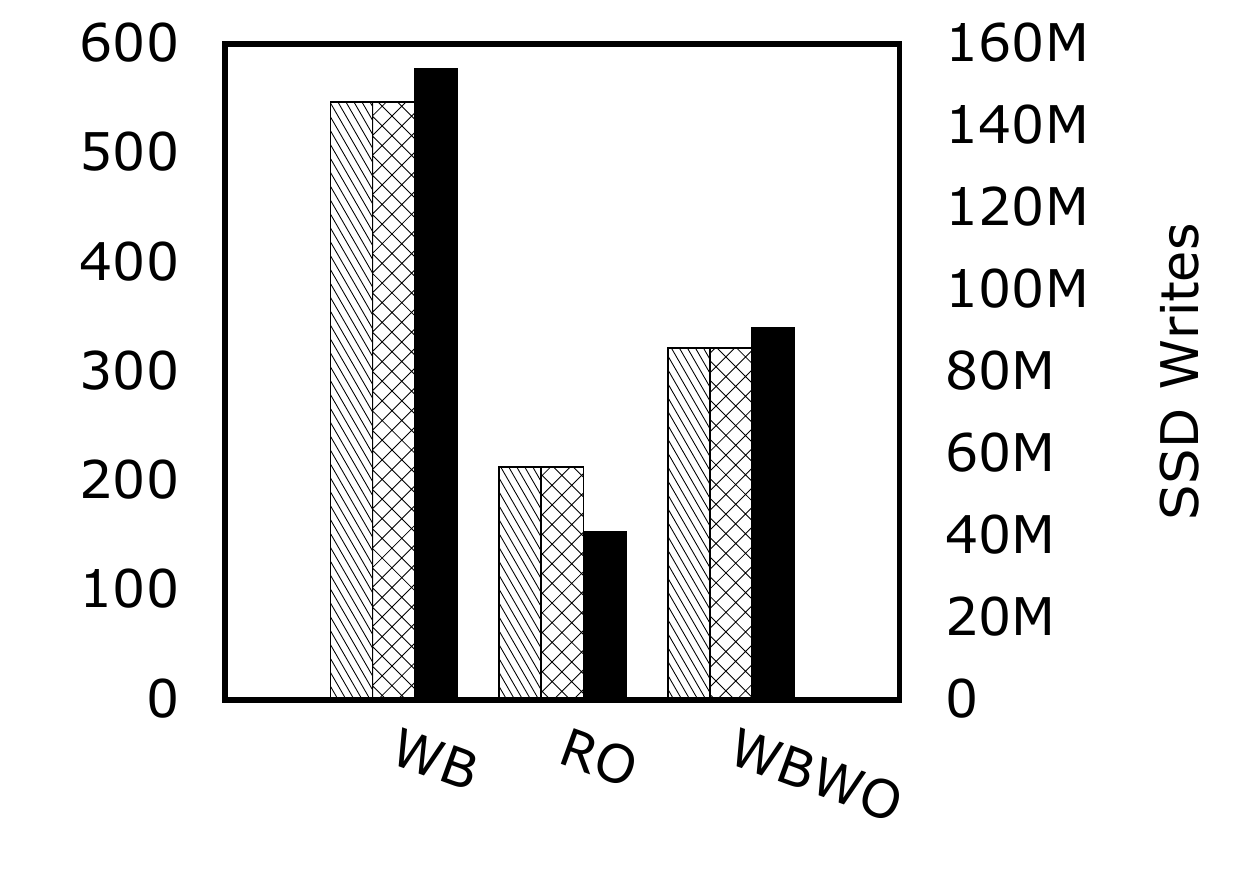}%
		\label{fig:varmail}}

	\caption{Effect of the cache write policy on the performance of workloads and endurance of the SSD cache (WB: Write Back, RO: Read Only, WBWO: Write Back and Write Only, and IOPS: Input/Output Operations Per Second).}
	\label{fig:simple_exp_policy}
\end{figure*}

Fig. \ref{fig:simple_exp_policy} shows the results of the {experiments, in terms of both}  performance {(measured as \emph{Input/Output Operations Per Second} (IOPS))} and endurance ({measured} in terms of committed write operations on the SSD).
We make four major observations:
\begin{enumerate}[leftmargin=*]
	\item In the FIO-RandRW workload, {WB} {achieves} the {highest} performance. {But, WB also} imposes the {highest number of} write operations on the {SSD} compared to other policies. As shown in Fig. \ref{fig:fio-randrw}, by using WBWO {cache,}  performance decreases by 10.7\% while the number of write operations on the SSD {also} decreases by 32\%. {The} RO cache provides about 78\% lower performance compared to {the} WB cache while it decreases the SSD‌ writes by 82\%. This is because the FIO-RandRW workload mainly includes {\emph{Read After Write} (RAW)} operations (70\% read operations{),} where buffering {the} write {requests} in the cache ({as in the} WB and WBWO policies) significantly improves both read and write performance. WBWO policy {does not cache} read misses and hence decreases {the number of} SSD‌ writes by 32\% with only 10\% performance degradation, {compared to the WB cache}. On the other hand, RO {does not help the performance of} RAW operations and, therefore, cannot provide {good} performance. In addition, RO {does not cache I/O} write {requests} and hence, provides no performance improvement for {them}.
	\item In the Web Server workload, the WB policy achieves {the highest} performance {while also having the highest number of} write operations on the {SSD}. {The} difference in performance provided by {the} three policies is very {low: Fig. \ref{fig:webserver} shows} that {the} WB cache achieves about 15\% higher performance compared to WBWO by imposing 98\% more writes on the {SSD}.
	{Compared} to RO, WB only marginally improves performance {(by 2.7\%)} at the cost of imposing 11.3\% {more} writes on the {SSD}. This is {because the} Web Server workload mainly includes random cold reads (i.e., the first read access to an address) with low locality of {reference,} and keeping such data blocks in the SSD‌ cache {does} not help in improving read performance. The WB and WO caches keep read miss data blocks in the SSD cache and impose about 178M and 158M writes on the SSDs, respectively, while WBWO{, compared to WB,} {does \emph{not} cache} read misses and achieves almost similar read performance (about 15\% lower) with {a} significantly {smaller} number of writes on the SSDs (about 98\% {lower}).
	\item In the Video Server workload, {all three polices} achieve the same read and write {performance} with considerable difference in {the number of} SSD writes. As shown in Fig. \ref{fig:videoserver}, WB, RO, and WBWO achieve 382 IOPS in read and 0 {IOPS} in write requests (the running workload  includes {\emph{only}} read requests).  WB and RO impose about 29M writes on the SSD {cache} while WBWO does {\emph{not}} {perform} any {write operations on}  the SSD cache. This is because the Video Server includes sequential read requests without any locality of {reference}. {Thus,} WB and RO {perform a} large number of unnecessary writes (due to read misses) into the SSD cache without any performance gain. WBWO imposes no writes to the SSD cache and achieves the same performance as WB and RO. Note that for such {a} sequential read workload, all requests are supplied from {the HDD} and the cache hit ratio is equal to 0. 
	\item In the Varmail workload, {WB has the highest} performance and SSD writes.
	{Both WBWO and RO reduce}  performance and SSD writes {over WB}.
	{This is because the RO and WBWO caches fail in serving RAW and RAR accesses, respectively.}
	Fig. \ref{fig:varmail} shows that both read and write performance {of} WB are about 41\% more than WBWO while WBWO imposes about 40\% {fewer} writes on the SSD. 
	{Compared} to RO, WB achieves 61\% {higher} performance by {imposing} 73\% more writes into the {SSD}. The reason is that Varmail includes {an} equal number of random read and write requests. WBWO {does not cache} read misses and hence cannot {help the} RAR requests, {but} it reduces the number of writes on the {SSD}.  RO does not buffer write requests and {thus} RAW {requests, but it reduces the number of writes on the SSD}.
\end{enumerate}
Our main experimental conclusions are:
\begin{enumerate}[leftmargin=*]
	\item In most of the workloads, WB cache achieves the {highest} performance for read and write requests, {however it also imposes the highest number of writes on the SSD}.
	{Hence, WB} does {\emph{not}} balance performance and endurance.
	\item In workloads with {a} large number of cold reads (such as Web Server and Video Server), WBWO has almost the same performance as WB while {it imposes a much smaller} number of writes on the SSD cache.
	\item In workloads with {a} large number of read requests (such as FIO-RandRW), WBWO cannot provide {as good} read {or} write performance {as WB}, {but} it results in a {smaller} number of SSD writes.
	\item RO cache can be employed in {read-intensive} workloads with {a} large number of re-references to {greatly} extend the SSD lifetime with negligible performance {impact compared to WB}.
	\item {In workloads such as Video Server and Web Server,} cache read misses impose {a} large number of write operations on the {SSD,} {affecting} both endurance and performance. 

\end{enumerate}

\section{\techname{} Architecture}
\label{sec:arch}
In this section, we present the architecture of {our} proposed two-level I/O cache, \techname{}. \techname{} {has four major characteristics. It:} 1) employs {both} DRAM and SSD in the caching layer of virtualized platforms, 2) assigns effective cache write policies to the {two} levels of {cache}, 3) effectively detects the popular data blocks and aims to evict only unpopular blocks {from the cache}, and 4) allocates efficient cache {space} for the VMs using the \emph{\URDPfull{}} (\URDP{}) metric.

\begin{figure}[!h]
	\centering
	\includegraphics[scale=0.9]{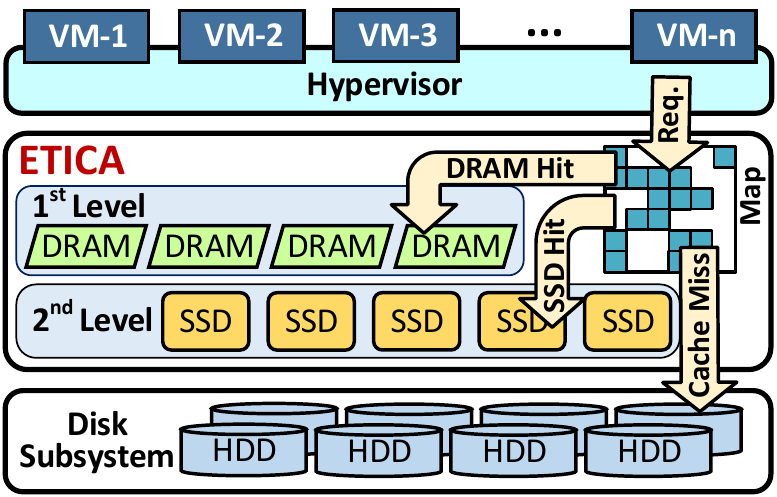}
	\caption{Architecture of \techname{}.}
	\label{fig:2-level-virtualized}
\end{figure}

Fig. \ref{fig:2-level-virtualized} provides an overview of the proposed two-level cache for virtualized platforms. 
{The} I/O cache employs two levels including DRAM in the first and the SSDs in the second level. The information about the content of cache levels {is} kept in a table {called} \emph{Map}. 
The hypervisor receives the I/O requests of the VMs, {and} routes them to the storage subsystem. The two-level cache resides {on} the path of requests to the disk subsystem. The cache searches the destination address of the requests in \emph{Map} and finds out whether the corresponding data is located in the $1^{st}$ level (i.e., DRAM hit) or  the $2^{nd}$ level (i.e., SSD hit). {If neither is the case}, the request is supplied by the disk subsystem (i.e., {cache} miss).\footnote{{Each cache level works based on set-associative mapping scheme where each set size is 512 blocks (the set size is configurable to other sizes, but to have a fair comparison, in our experiments the associativity configuration in ETICA is done the same as ECI-Cache).}}

We aim to take advantage {of the different} merits of {both the} DRAM technology and SSDs while minimizing the negative {characteristics} of each.
DRAM {provides} high performance but {comes} with 1) reliability {issues (due to volatility)} and 2) high cost. {SSDs preserve reliability {(due to non-volatility)} {and have} 20X lower cost than DRAM, but they have 1) lower performance and 2) limited endurance (i.e., write cycles)}.
As shown in Fig. \ref{fig:2-level-virtualized}, DRAM in the first level of the proposed cache mainly improves performance {(and endurance)} while the SSD in the second level avoids the performance spikes due to {the} large performance gap between DRAM and {the} disk subsystem {and provides reliability}. In Section \ref{sec:write_policy}, we describe how we assign {different} write {policies}  {at different} levels of cache to  address {the shortcomings and exploit the advantages of each technology}.
{In} Section \ref{sec:promote_evict}{, we} propose popular/unpopular block detection and our promotion/eviction schemes {to manage the two-level cache even more efficiently}.
{In} Section \ref{sec:size_estimation}, we describe {how \techname{} estimates} {efficient} cache {sizes} for {different} VMs.

\subsection{Write Policy Management}
\label{sec:write_policy}
We use {DRAM} in the first level of {our} I/O cache{ architecture {to enable high performance}. Using}  such {a} volatile storage {technology}  increases the risk of data loss under different {types of} failures, such as power {failures.} {We use SSDs} in the second level of our I/O cache architecture. {SSDs} mainly suffer from endurance{, i.e., they can support only a} limited number of reliable writes. 
This section shows how we 1) preserve {storage} reliability {in the presence of DRAM} and 2) enhance {SSD} endurance in our proposed two-level I/O cache by applying effective write policies on both levels.
Previously, in Section \ref{sec:motivation}, we described the different types of cache write policies and their impact on performance, reliability, and endurance. 

To address {the storage} reliability {issue with using DRAM}, we assign {the} RO policy to the first level of cache (i.e., {the} DRAM level). In this case, {\emph{no}} write requests are {served (i.e., supplied)} by DRAM and hence, there is no write-pending request (i.e., dirty block) in DRAM. Write requests are directed {immediately} to the {non-volatile} second level (i.e., {the} SSD level{),} which is able to protect {the write-pending} data in case of power outage (it is important to note that in our scheme, we use SSDs in {the} RAID10 configuration \cite{chen1994raid}{, which} tolerates {the failure of} up to two SSDs). In this scheme, {the} DRAM level is responsible {for} buffering and serving {only} read requests. 
{Buffering} only read {requests} in DRAM {guarantees that} any data {block} in {the} DRAM level {has} a copy in {another}  level ({i.e.,} {the} SSD level or {the} disk subsystem). {Thus,} losing data in DRAM has no negative reliability impact.

{To address the endurance issue of the SSDs, we assign the WBWO policy to the second level (i.e., the SSD level). Recall that in} Section \ref{sec:motivation}, we observed the significant {negative} impact of buffering read misses on {the} endurance of the SSD cache. {The} {WBWO} {policy buffers} {only} write requests {in the cache} and {does {\emph{not}} buffer} read {requests}.
{Buffering only write requests in the SSDs} effectively improves the endurance of {the} SSDs by reducing {the} number of {unnecessary} write {operations on the SSDs} ({that would otherwise be needed to buffer read requests}).
{Using WBWO at the SSD level does not degrade performance because read requests are buffered and served at the DRAM level.}

\begin{figure}[!htb]
	\centering
	\subfloat[One-level SSD Cache]{\includegraphics[width=.5\textwidth]{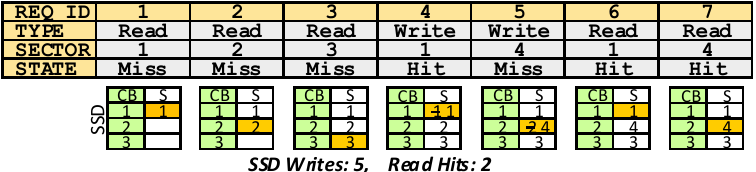}%
		\label{fig:motivation_exmp_1-level}}
	\hfil
	\subfloat[Proposed two-level DRAM+SSD cache]{\includegraphics[width=.5\textwidth]{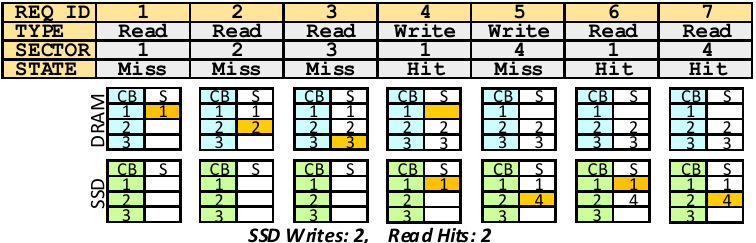}%
		\label{fig:motivation_exmp_2-level}}
	
	\caption{Comparison of a) one-level and b) the proposed two-level cache in terms of endurance, reliability, and performance (CB: Cache Block and S: Sector).}
	\label{fig:mot_example}
\end{figure}

In summary, the key advantages of {our heterogeneous}  write policy assignment {for two different cache levels} are:
\begin{enumerate}[leftmargin=*]
	\item We preserve {the} reliability of write requests by bypassing DRAM and buffering them only in SSDs (in RAID10 configuration).
	\item We improve {the} endurance of the SSDs by buffering only write requests while read misses (which {would} impose {a} large number of write {operations on the SSDs}) are only cached in {the} DRAM level.
	\item The {DRAM} level {improves the performance of {read} requests} while {the} {SSD} level {improves the} performance of  {write} requests{, compared to HDDs}. {The} read misses from DRAM {that} {happen to be} supplied by {the} SSDs ({due to} RAW {requests}) {also} experience the {higher} performance of SSDs {versus the} disk subsystem.
\end{enumerate}
We now provide an example and show how the proposed {write} policy assignment approach in our two-level cache 1) improves {the} endurance of SSDs, 2) preserves the reliability of I/O requests, and 3) keeps performance intact. In the example of Fig. \ref{fig:mot_example}, we compare {a} one-level WB SSD cache (Fig. \ref{fig:motivation_exmp_1-level}) with our two-level cache where the write policy of {the} DRAM and SSD levels are RO and WBWO, respectively (Fig. \ref{fig:motivation_exmp_2-level}). {Each {cache level has a capacity of 3 data pages.}} Note that in this example, no cache size estimation algorithm is employed.

{When one-level SSD caching is used,} as shown in Fig. \ref{fig:motivation_exmp_1-level}, $Req_1$ reads the data of sector 1 and  {buffers} the corresponding data block {into} the cache (due to read miss) {by} imposing one write operation {on} the SSD. Similarly, $Req_2$ and $Req_3$ are read misses {that} are {buffered in} the cache and each one imposes {an} additional write {on} the SSD. 
$Req_4$ overwrites the {data} content of {the} cache (sector 1 in cache block 1) and $Req_5$ writes the contents of sector 4 into the cache (in cache block 2). $Req_6$ hits in the cache and reads sector 1 from cache block {1,} which was previously written by $Req_4$. Similarly, $Req_7$ reads data from cache as {it is a} read hit. We observe that, in {the} one-level WB SSD cache, the sample workload imposes 5 write operations to the SSD and {obtains} 2 read hits.

Fig. \ref{fig:motivation_exmp_2-level} shows how {the two-level cache design of} \techname{} (with RO DRAM and WBWO SSD levels) supplies the I/O requests with {a much smaller number of} write operations on the SSD.
In the two-level cache, $Req_1$, $Req_2$, and $Req_3$ {buffer} data blocks {into} the DRAM level (due to read misses) without imposing {any} write operations {on} the SSD. {$Req_4$} overwrites the cache content (sector 1{):} the write request is served by {the} SSD and the existing data block in DRAM is marked as invalid. $Req_5$ is a write request {that} bypasses DRAM and writes data {directly} {into} the SSD. Finally, $Req_6$ and $Req_7$  are read requests {that} hit in the cache and read the data blocks {from the} SSD level.
Unlike {the} one-level SSD cache, our two-level cache {design} imposes only 2 write operations {on} the SSD (60\% {fewer} than one-level) while providing the {same} hit ratio as one-level does.
In summary, compared to {the} one-level {SSD} cache, our two-level {DRAM+SSD} cache {with {intelligent} cache write policy assignment} reduces the number of write {operations} {on} the SSD while {preserving} the reliability of write requests and {achieving} similar performance.\footnote{{Performance of our scheme would be higher when requests hit in the DRAM level.}}

\subsection{Promotion/Eviction Optimization}
\label{sec:promote_evict}

In this section, we propose another approach for improving both endurance and performance in our two-level I/O cache, in which we mainly focus on improving  performance (in terms of hit ratio) with the minimum {number of write operations} into the SSD.
The proposed approach {is mainly} adopted from \emph{pull mode caches} (i.e., \emph{non-datapath caches}), where the content of the SSD cache is not updated during handling of miss accesses (as opposed to \emph{push mode caches} or \emph{datapath caches}) \cite{wec2015}.
Updating cache blocks on each cache miss (as \emph{push mode {caches}} such as LRU\footnote{Least Recently Used} and LFU\footnote{Least Frequently Used} {caches} do) leads to promoting new blocks to the cache which {could easily be} less popular than the evicted ones.
Instead, periodically, we detect \emph{unpopular} data blocks and evict them from {the} SSD and promote \emph{popular} ones.
{Thus}, we minimize the probability of evicting \emph{popular} blocks from {the} cache, thereby {avoiding}  performance (due to cache miss) and endurance (due to promoting {new} data blocks into the cache) overheads {associated with the churn caused by the pollution due to unpopular blocks}. 
To do so, we first detect \emph{popular} and \emph{unpopular} data blocks and then decide \emph{when} to promote and evict them to/from the SSD level.

We aim to detect \emph{popular} and \emph{unpopular} data blocks {via} online characterization of the running workload  on the VMs.
To this end, in specific time intervals (which are determined based on the number of requests, e.g., after {observing} N {I/O} requests), we analyze the characteristics of the  running workloads based on the collected information such as 1) request type, 2) number of accesses, and 3) \URDP{} of data blocks.
In Section \ref{sec:popular_detection}{, we} {describe} how we detect \emph{popular} and \emph{unpopular} data blocks.

\begin{figure}[!h]
	\centering
	\subfloat[Read Requests]{\includegraphics[width=.23\textwidth]{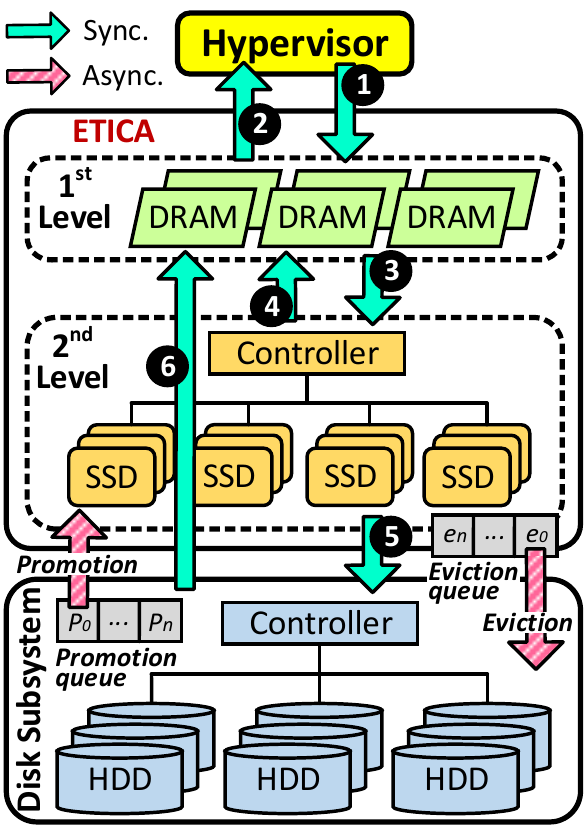}%
		\label{fig:2-level-read}}
	\hfil
	\hspace{-1pt}
	\subfloat[Write Requests]{\includegraphics[width=.23\textwidth]{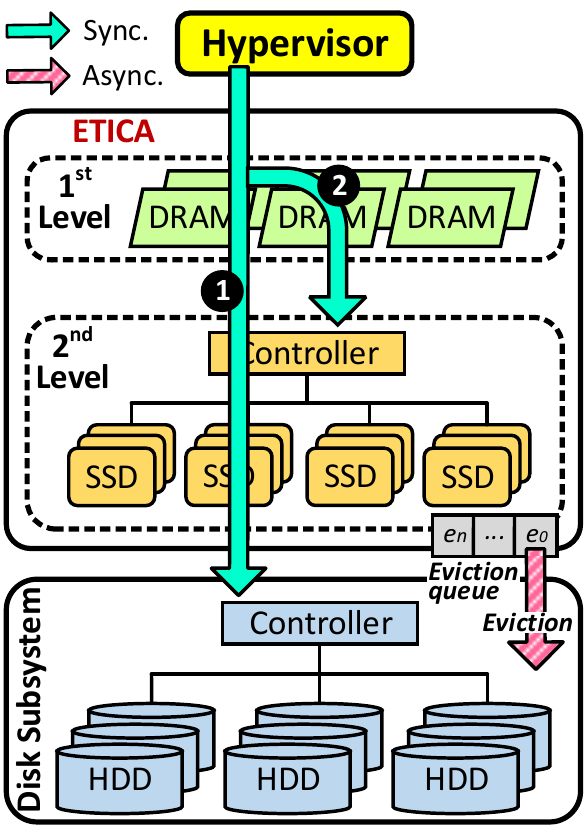}%
		\label{fig:2-level-write}}
	
	\caption{{Examples showing how \techname{} handles requests.}}
	\label{fig:2-level-arch}
\end{figure}

Fig. \ref{fig:2-level-arch} shows the flow of {I/O} requests in our proposed two-level I/O cache for both read and write accesses.
As shown in this figure, we employ two queues in our scheme: 1) {\emph{promotion queue}} in {the} disk subsystem level and 2) {\emph{eviction queue}} in {the} SSD cache level.
The data blocks in the disk subsystem {that} are recognized as \emph{popular} are pushed into the promotion queue while the \emph{unpopular} ones in the SSD cache are {inserted into} the eviction queue.
Periodically, we evict the blocks in {the} eviction queue from {the} SSD‌ cache to the disk subsystem and promote the blocks in {the} promotion queue {into} the SSD‌ cache (only when {there is} free space in SSD).
{This} approach {greatly increases the likelihood} that 1) data blocks are evicted from {the} SSD when they become \emph{unpopular}, 2) only \emph{popular} data blocks are buffered in the SSD cache, and 3) {we do not replace}  a data block in the SSD cache with a less popular one{. Hence, the proposal improves both}   performance and SSD‌ endurance.

We now elaborate on  how {our} proposed promotion and eviction approach in our two-level I/O cache handles the {I/O} requests. ‌Fig. \ref{fig:2-level-read} and Fig. \ref{fig:2-level-write} provide the detailed flow of read and write requests, respectively.  
\paragraph{\textbf{Read}} 
As shown in Fig. \ref{fig:2-level-read}, the read request is {first} received in {the} DRAM level (\circled{1}). If the corresponding data block is found in {the} DRAM cache (i.e., DRAM hit), the request is served via {the DRAM} level (\circled{2}). Otherwise, the request is sent to the SSD‌ in the second level (\circled{3}). In {the} case of {an} SSD hit, we first  promote the data block to {the} DRAM level (\circled{4}) and then serve the request. Otherwise, in case of {an} SSD‌ miss (\circled{5}) we read the data from {the} disk subsystem and promote it directly to the DRAM level {\emph{without}} buffering in {the} SSD‌ level (\circled{6}). Recall {that} the write policy management scheme presented in Section \ref{sec:write_policy} {sets} the policy of {the} SSD level to {WBWO,} where no read miss is buffered in {the} SSD. However, any data block {that} is detected as \emph{popular} in the disk subsystem (listed in {the} promotion queue) is promoted to the SSD level to accelerate future accesses. Furthermore, \emph{unpopular} {data blocks in the SSD} are evicted  through {the} eviction queue.

\paragraph{\textbf{Write}}
Fig. \ref{fig:2-level-write} shows how we handle write requests in the proposed two-level I/O cache.
Write requests (in case of both hit or miss) bypass the DRAM (since the write policy of {the} first level is set to RO).
In case of {an} SSD‌ miss (\circled{1}), the data block is directly written {to} the disk subsystem.
Otherwise, in {the} case of {an} SSD‌ hit {where} the corresponding data block is located in the second level (as a \emph{popular} block), the data is written {to} the SSD level (\circled{2}).
{Note that in case of data block update in any level (SSD and disk subsystem), we invalidate the corresponding data block in upper level (DRAM and SSD, respectively).}
The operation of {the} eviction queue in handling writes is the same as {that of the} eviction queue in handling read requests.

To summarize, the proposed eviction/promotion scheme provides the following key benefits:
\begin{enumerate}[leftmargin=*]
	\item Performance improvement by keeping \emph{popular} data blocks in the {SSD} cache as long as possible.
	\item SSD endurance improvement (reduced number of SSD cache updates) by periodically promoting \emph{only} popular data blocks into the {SSD} cache.
\end{enumerate} 

  \subsubsection{Popular and Unpopular Block Detection}
\label{sec:popular_detection}
In this section, we provide our approach {to} detecting \emph{popular} and \emph{unpopular} data blocks.
Our proposal decides the popularity of data blocks based on two key parameters: 1) \URDP{} and 2) frequency of accesses{. H}ence, {we} consider both \emph{spatial} and \emph{temporal} {locality}.
We update the $popularity(B_i)$ parameter for each access to the $i^{th}$ data block $B_i$ and push the data block into the eviction or promotion queues based on {its} relative {popularity}. The {least popular} 5\% {of the blocks} in the SSD (i.e., the \emph{unpopular ones}) are put into the eviction queue. Similarly, the {most popular} 5\% of {the} data blocks in {disk subsystem (i.e., HDD)} are pushed into the promotion queue.
Eq. \ref{equ:popularity} shows how we calculate $popularity(B_i)$:
\begin{equation}
popularity(B_i) = \sum_{t=1}^{numAcc}e^{-{POD(i,t) \over cacheSize}}
\label{equ:popularity}
\end{equation}

\begin{figure}[!h]
	\centering
	\includegraphics[scale=.26]{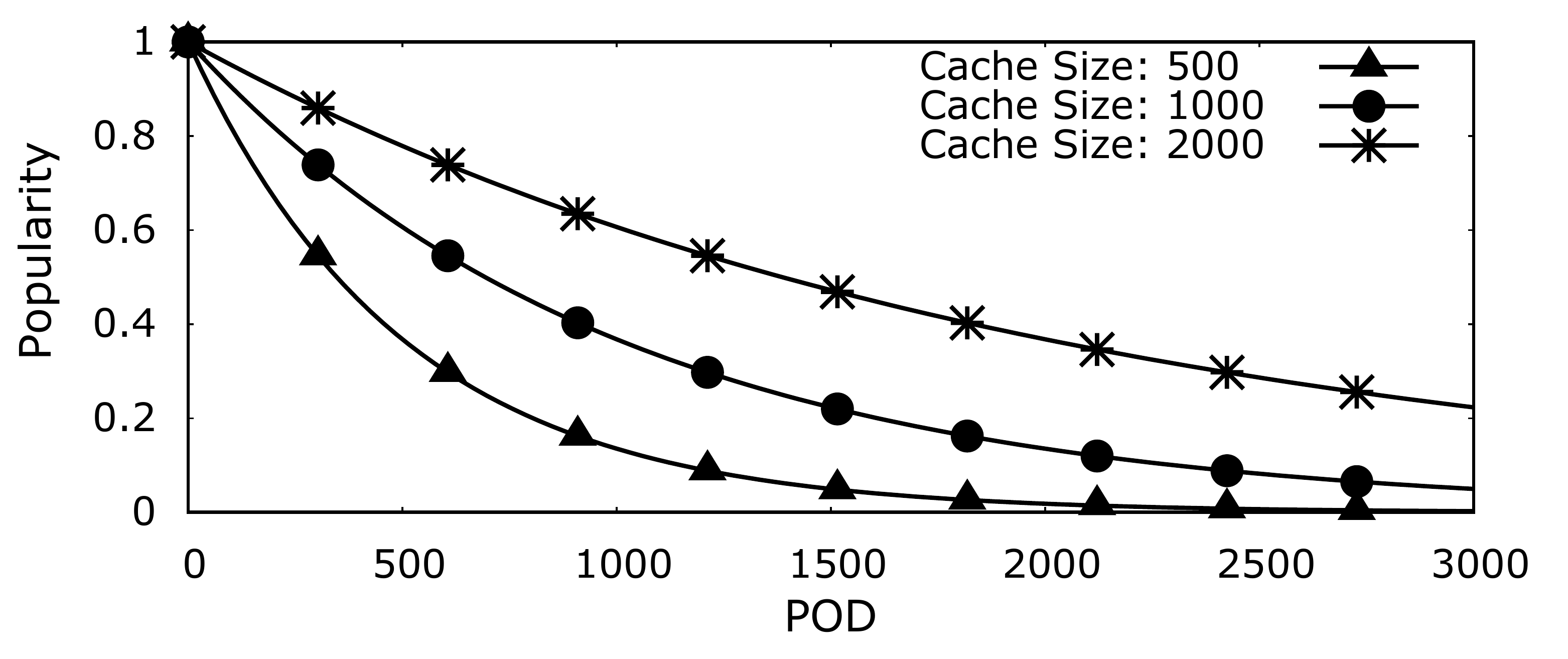}
	\caption{{Popularity as a function of \URDP{} and cache size.}}
	\label{fig:popular-chart}
\end{figure}

Where $POD(i,t)$ is the \URDP{} of $B_i$ in the $t^{th}$ access to that data block,\footnote{In Section \ref{sec:urd+} we elaborate how to calculate \URDP{}.} $numAcc$ is the number of accesses to $B_i$, and $cacheSize$ is the allocated cache size to the VM.
{Popularity of each data block is updated periodically in specified time intervals. To keep the popularity information of each data block, a space of 8B is required per data page (4KB). {The memory overhead is} less than 0.15\% (e.g., we need {an} 8MB space to store the popularity of one million data blocks. {This} information {is} kept in SSD). Since popularity is computed asynchronously, its computation does not {cause} any overhead to {the} servicing {of} I/O requests.}
Fig. \ref{fig:popular-chart} shows the impact of \URDP{} on {the} $popularity$ of each access for different cache sizes.
It can be seen that  calculating $popularity(B_i)$ based on Eq. \ref{equ:popularity} provides the following key {ideas}:
\begin{enumerate}[leftmargin=*]
	\item {A larger} \URDP{} leads to smaller \emph{popularity}. Note that when the \URDP{} of {a block} is close to the cache size, the frequency of  access {to that block} is low. {Thus, that block likely} has no significant impact on {the} total {cache} hit ratio.
	\item When the \URDP{} of {a} {block} is larger than {the} cache size, {and} since this {block} will be missed from cache, we set low \emph{popularity} to that {block} according to Eq. \ref{equ:popularity}. On the other hand, we set high \emph{popularity} for {a} {block} with {a} \URDP{} smaller than {the} cache size.
	\item {The frequency of the accesses is considered in {the} popularity calculation. In Eq. \ref{equ:popularity}, we estimate the total popularity of a single block based on all accesses to that block.}
\end{enumerate}

The proposed popularity detection scheme {is scalable, which is compatible with {the} removal or addition of VMs {into the system}.
	{When we add} a new VM to the system (or when an existing VM {comes} online), based on the I/O accesses of that VM, \techname{} first assigns efficient cache space in both DRAM and SSD levels and then estimates the popularity of that VM's accesses (we calculate popularity of each {VM's accesses independently from other VMs}). 
	In contrast, when a VM is removed from the system (or when it {goes} offline), we reclaim the allocated cache space and thus the popularity detection process for this VM {is also} stopped. Furthermore, since in specific intervals, \techname{} recalculates the cache size of the VMs, in case of extending or shrinking the total SSD and DRAM cache spaces, \techname{} is able to reconfigure the newly allocated cache sizes to the VMs based on new SSD‌ and DRAM sizes.}

\subsection{Cache Size Estimation}
\label{sec:size_estimation}

In this section, we propose our two-level cache size estimation approach {that} aims to {1)} maximize the overall performance of the {co-running} VMs {and 2) minimizes the overall cost} (in terms of {allocated} cache size).
We effectively partition the space of both levels (DRAM and SSD) between VMs based on their demand.
In {Section} \ref{sec:urd+}{, we} propose the metric of \emph{\URDPfull{}} (\URDP{}) and demonstrate how the proposed metric effectively reduces the allocated cache size to the workloads (compared to {the} previous state-of-the-art scheme, URD) by considering both 1) \emph{{request} type} and {2)} \emph{cache write policy} in {the} reuse distance calculation, {and thereby} resulting in reduced cost. 
Then in Section \ref{sec:eff_size} we show how \techname{} assigns efficient cache size for the VMs to achieve the most efficient performance per cost.

\subsubsection{\URDPfull{} (\URDP{})}
\label{sec:urd+}
\emph{Traditional Reuse Distance} (TRD) \cite{mattson, parda, shards,ding2003predicting,zhong2009program,fang2004reuse,ding2001reuse,berg2004statcache,loc-appr-conf,Zhao:RD-1,Wu:2013:ERD,Badamo:RD-2} {calculates} the distance of the requests \emph{only} based on the address {and access order} of the requests. \emph{Useful Reuse Distance} (URD) \cite{ahmadian2018eci} improves {upon} TRD by considering \emph{request type} in reuse distance calculation{: it} considers {\emph{only}} the distance of \emph{Read After Read} (RAR) and \emph{Read After Write} (RAW) {accesses, enabling a smaller cache with the same {or with better} performance} \cite{ahmadian2018eci}.
However, existing schemes {that} are employed in cache size estimation neglect the impact of \emph{cache write policy} on reuse distance calculation. In this section, we first provide examples to show the {effect} of {the} \emph{cache write policy} on reuse distance calculation of the workloads. Then, we propose {a novel metric,}  \emph{\URDPfull{}} {(\URDP{}),} which allocates much smaller cache space compared to TRD‌ and URD while preserving performance (in terms of hit ratio). 
{The} {\URDP{} metric considers both \emph{request type} and \emph{cache write policy} in reuse distance calculation, and hence, does {\emph{not}} reserve cache blocks for the requests which would {\emph{not}} be served by the cache (i.e., read accesses in a WBWO and write accesses in a RO cache, respectively).}
We provide sample workloads and elaborate {on} how URD (which considers only the request type without considering {the} cache write policy) and \URDP{} (which considers {\emph{both}} {request} type and {the} cache write policy) estimate cache size for caches {that use} 1) WBWO and 2) RO write policies.

\begin{figure}[!htb]
	\centering
	\subfloat[]{\includegraphics[height=0.18\textwidth, width=.265\textwidth]{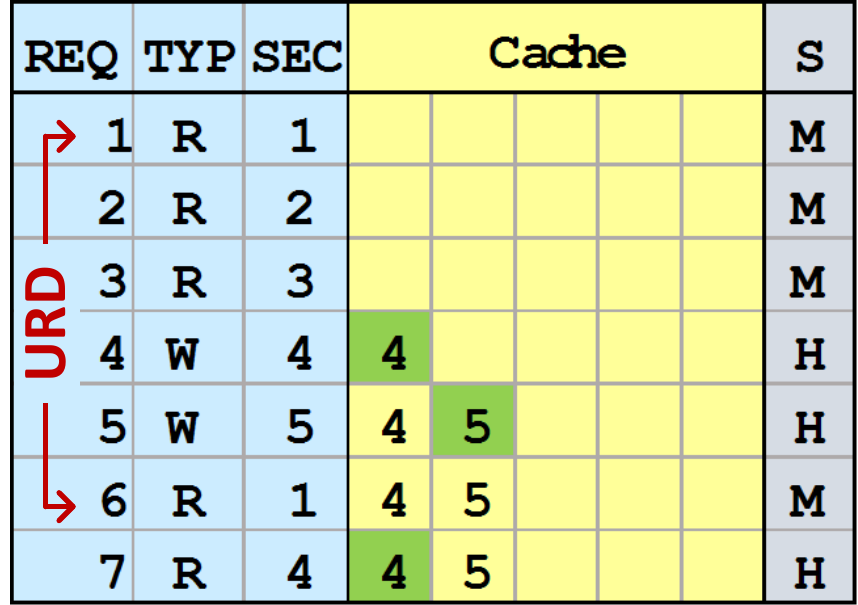}%
		\label{fig:urd_wbwo}}
	\hfil
	\subfloat[]{\includegraphics[height=0.18\textwidth, width=.19\textwidth]{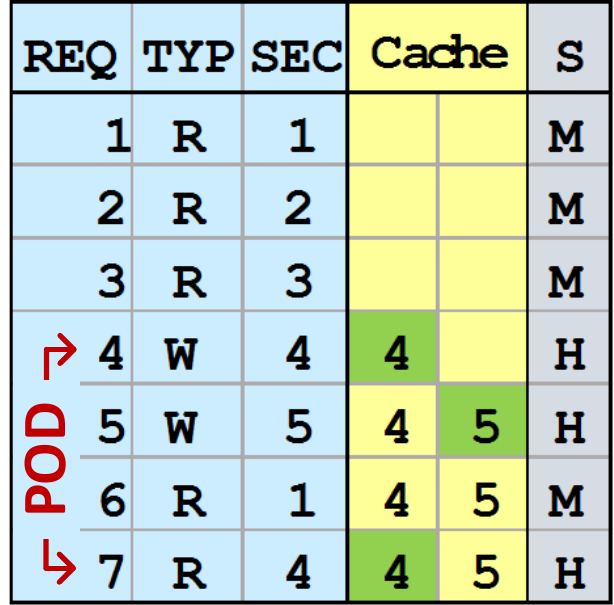}%
		\label{fig:urd+_wbwo}}
	
	\caption{Comparison of cache size allocation in {a} WBWO cache (a) without  and (b) with considering cache write policy (REQ: Request, TYP: Type, SEC: Sector, W: Write, R: Read, S: State, H: Hit, M: Miss).}
	\label{fig:example_wbwo}
\end{figure}

Fig. \ref{fig:example_wbwo} compares the cache size estimation {provided} by URD and \URDP{} in a WBWO cache.
For the given sample workload in Fig. \ref{fig:urd_wbwo}, URD detects the \emph{maximum} reuse distance between $Req_1$ and $Req_6$ {(due to their RAR accesses to the same sector, i.e., sector 1).} Hence, the maximum URD is equal to {4, and} 5 blocks of cache are allocated to the workload. 
Fig. \ref{fig:urd_wbwo} {shows} that the workload uses only two blocks of the allocated cache space for write requests (write accesses to sector 4 and sector 5 by $Req_4$ and $Req_5$, respectively). Since data blocks of \emph{read misses} are {\emph{not}} buffered in a WBWO cache, three blocks of allocated cache remain {\emph{unused}} while they are  reserved for this workload. At the end of {the} workload, $Req_7$ reads the buffered data block from {the cache: it}  is a read hit due to {the} RAW operation.
We observe that a cache with WBWO policy {helps only} the write and RAW\footnote{Note that WBWO cache supplies \emph{only} RAR accesses where the first read was RAW (i.e., RARAW). We also consider such accesses as RAW and calculate \URDP{} for the \emph{maximum} distance of such accesses.} {requests} without {providing} any performance improvement for read and RAR (i.e., Read After \emph{Cold}\footnote{The first accesses to an address.} Read) accesses.

The \URDP{} metric considers both \emph{request type} (similar to URD) and \emph{cache write policy} in reuse distance and cache size estimation.
Fig. \ref{fig:urd+_wbwo} shows how \URDP{} estimates cache size for the sample workload. \URDP{} estimates the \emph{maximum} reuse distance based on $Req_4$ and $Req_7$ due to their RAW access to sector 4. In this case, the \URDP{} of the workload is equal to 1.
{Crucially, }note that the read access of $Req_6$ is {\emph{not}} considered in \URDP{} calculation in the WBWO cache. This is because such {a} request would {\emph{not}} be served by the cache. Therefore, {only} two blocks of cache are allocated to the workload. $Req_4$ and $Req_5$ commit write requests into the cache (due to their accesses to sectors 4 and 5) and $Req_7$  {(to sector 4)} is supplied from {the} cache.
In summary, for the given sample workload, we observe that in a WBWO cache URD allocates 5 blocks while \URDP{} allocates only 2 blocks of cache and achieves the same I/O performance (for both read and write {requests}) as URD.

\begin{figure}[!htb]
	\centering
	\subfloat[]{\includegraphics[height=0.18\textwidth, width=.265\textwidth]{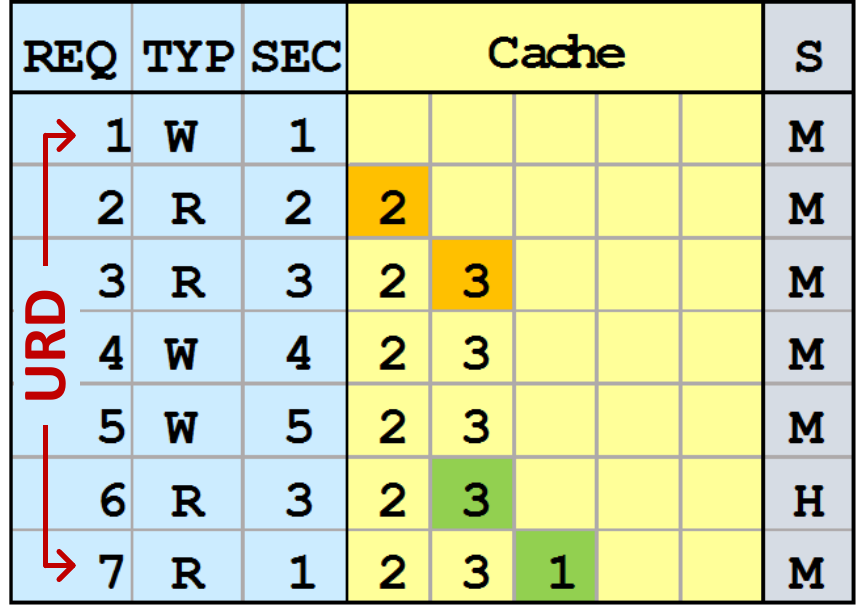}%
		\label{fig:urd_ro}}
	\hfil
	\subfloat[]{\includegraphics[height=0.18\textwidth, width=.19\textwidth]{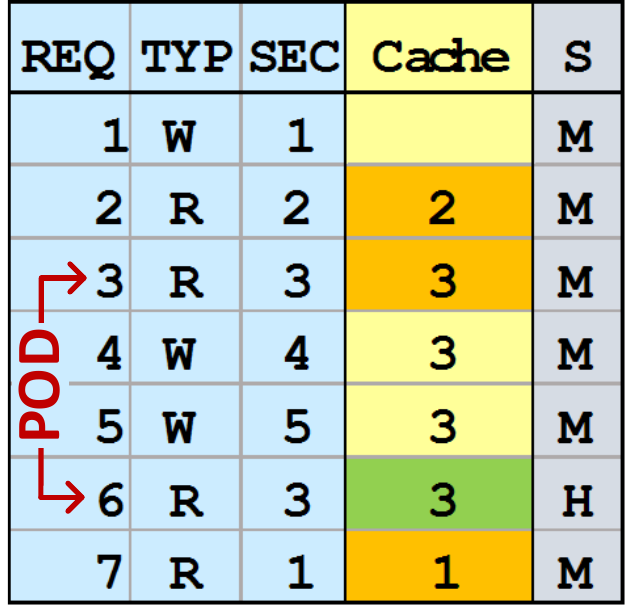}%
		\label{fig:urd+_ro}}
	
	\caption{Comparison of cache size allocation in RO cache (a) without  and (b) with considering cache write policy  (REQ: Request, TYP: Type, SEC: Sector, W: Write, R: Read, S: State, H: Hit, and M: Miss).}
	\label{fig:example_ro}
\end{figure}

Fig. \ref{fig:example_ro} shows how URD and \URDP{} work in a RO cache{, by comparing} {the} cache {allocation} and I/O performance {of} these {two} schemes. For the given sample workload in Fig. \ref{fig:urd_ro}, URD detects the maximum reuse distance for the RAW access by $Req_1$ and $Req_7$ to sector 1 of the disk. In this case, the URD of the workload is equal to 4 and hence, five blocks of cache are allocated to the workload (note that read access of $Req_6$ to sector 3 is duplicated with $Req_3$ {and} both of them are supplied by {the same} {cache} block).
{Fig. \ref{fig:urd_ro} shows that the}  write accesses of the workload ($Req_1$, $Req_4$, and $Req_5$) are bypassed from {the} cache and are supplied by {the} disk subsystem. Only the \emph{read misses} ($Req_2$ and $Req_3$) are buffered in cache and RAR accesses such as $Req_6$ are served {while RAW accesses such as $Req_7$ cannot be served by the RO cache}.
{Only} three blocks of allocated cache are {actually} used and the remaining two blocks are unused {but} they are reserved by the workload. This is because the RO cache only buffers reads and serves RAR accesses.
In this case, reserving cache blocks for write accesses (as URD does) has no performance {benefit} while imposing {additional} cost ({because it allocates} larger cache space with unused cache blocks).

In Fig. \ref{fig:urd+_ro}{, we} show how \URDP{} considers the \emph{cache write policy} in reuse distance calculation and results in 1) {a} much smaller cache {space allocation} compared to URD and 2) the same I/O performance {as URD}.
In the RO cache, \URDP{} detects the maximum reuse distance of the workload for {the} RAR {accesses} to sector 3 by $Req_3$ and $Req_6$. Since no write request {can be} supplied by RO cache,  write requests are {\emph{not}} considered in {the} \URDP{} calculation of such {a} cache. Hence, \URDP{} is equal to 0 ($Req_4$ and $Req_5$ are not considered since write accesses have no impact on {the} RO cache operation) and {\emph{only one block}} of cache is allocated to the workload. As shown in Fig. \ref{fig:urd+_ro}, write accesses  ($Req_1$, $Req_4$, and $Req_5$) are not served by the cache. A read miss by $Req_2$ promotes {the} data block of sector 2 into the cache. Then, $Req_3$ updates the cache by promoting {the} data block of sector 3 (note that in this workload there is no future read access to sector 2 and \URDP{} enables updating the data block by $Req_3$). $Req_6$ hits in the cache and finally, $Req_7$ promotes sector 1 into the cache.
We observe that in {an} RO cache, \URDP{} calculates the maximum reuse distance based on RAR operations. Compared to URD, \URDP{} assigns {a} {much} smaller cache {space} to the workload {(only 1 block versus 5 blocks)} while achieving {exactly} the same {cache} hit ratio. 

To summarize, the key ideas behind \URDP{} are as follows:
\begin{enumerate}[leftmargin=*]
	\item \URDP{} considers both 1) \emph{request type} and 2) {the specific} \emph{cache write policy} in {the} reuse distance calculation.
	
	\item In a WBWO cache, {the} reuse distance calculation {of} \URDP{} is  based {only} on {the} \emph{maximum} distance of RAW accesses (also including RARAWs). This is because no RAR (i.e., Read After \emph{Cold} Read) {request} is supplied by {a} WBWO cache ({since} {the} cache {does {\emph{not}} buffer} read misses). {Hence,} \URDP{} does {\emph{not}} reserve cache blocks for {read miss requests} (note that the read hits are due to RAW {accesses} where the {read} cache block is previously {buffered} by {a} write {request}). {Thus, \URDP{}} {always} estimates {a} smaller cache size compared to URD{, while achieving}  the {\emph{same}} hit ratio and I/O performance (for both read and write operations {in a WBWO cache}).
	
	\item In {an} RO cache, {\URDP{} considers} the \emph{maximum} distance of {only} RAR accesses in reuse distance calculation. Since {the} cache {does {\emph{not}} buffer} write requests, no cache block is reserved for write operations by \URDP{}. In this case, \URDP{} results in {a} smaller cache size {than URD, while providing} the same hit ratio and I/O performance.
	
	\item In a WB cache, URD and \URDP{} work similarly. Both RAR and RAW accesses are supplied by {the} WB cache. In this case, {both URD and \URDP{} estimate} the cache size based on the \emph{maximum} distance of {either} RAR and RAW accesses. 
\end{enumerate}

\subsubsection{Allocating Efficient Cache Size for VMs}
\label{sec:eff_size}

{We now} elaborate on how we allocate {an} efficient cache size (in both DRAM and SSD levels) for {each} running {VM}.
\techname{} estimates and assigns cache sizes to the VMs by using the \URDP{} metric. \techname{} aims to achieve the maximum performance (in terms of hit ratio) with the minimum cost (in terms of cache size) as opposed to previous architectures which mainly aim to optimize performance with the maximum available cache sizes \cite{ahmadian2018eci,centaur}.
The steps of cache size assignment by \techname{} are as follows:
\begin{enumerate}[leftmargin=*]
	\item Periodically, {\techname{}} calculates the \URDP{} of the running VMs for both {the} RO and WBWO policies {used in the} DRAM and SSD levels, respectively. Recall from Section \ref{sec:urd+} that \URDP{} does {\emph{not}} reserve cache blocks for writes in RO and reads in WBWO caches and hence estimates different cache sizes for different write polices.
	
	\item Based on {the} calculated \URDP{}, we estimate {two} efficient cache {sizes} for the VM in both {the} DRAM and {the} SSD levels. {Eq. \ref{equ:size_est} shows how we calculate the cache sizes:}
	\begin{equation} 
	\begin{array}{l}
	Cache_{DRAM}=POD_{DRAM} \times blockSize_{DRAM}\\
	Cache_{SSD}=POD_{SSD} \times blockSize_{SSD}
	\label{equ:size_est}
	\end{array}
	\end{equation}	
	{Where $POD_{DRAM}$ and $POD_{SSD}$ are {the calculated} \URDP{} {metrics} of the running VM in DRAM and SSD levels and
	$blockSize_{DRAM}$ and $blockSize_{SSD}$ are the cache block size in DRAM and SSD{, respectively}.}
	
	\item We check whether the {total} estimated cache sizes {for all VMs} fit in the {available physical} DRAM and SSD space.
	
	\item If the sum of estimated {cache} sizes exceeds the existing {physical} DRAM or SSD {space}, we reduce the estimated size of each VM to maximize the \emph{Performance-Per-Cost} (PPC) factor in Eq. \ref{equ:conditions}.
	\begin{equation} 
	PPC=\sum_{i=1}^{numVMs}{H(VM_i,c_i) \over c_i}
	\label{equ:conditions}
	\end{equation}
	Where $numVMs$ is the number of running VMs and $H(VM_i,c_i)$ is the hit ratio {achieved} by $VM_i$ by allocating {it a} cache size equal to $c_i$.
\end{enumerate}

Since the running VMs are weighted identically, we aim to maximize the {aggregate} \emph{PPC} as proposed in Eq. \ref{equ:conditions}. Assigning cache sizes based on Eq. \ref{equ:conditions} guarantees the maximum overall performance with minimum cost. To do so, \techname{} provides \URDP{}-based \emph{Miss Ratio Curves} (MRCs) for each running VM and finds the set of $\{c_0, ..., c_{N-1}\}$ which provides the maximum \emph{PPC}.

\section{Experimental Results}
\label{sec:results}
In this section, we evaluate the effectiveness of our proposed two-level I/O {caching} scheme{, \techname{},} in terms of cache size, performance, and endurance. 
\begin{figure*}[!b]
	\centering
	\subfloat[VM0: \hm{}]{\includegraphics[width=.24\textwidth]{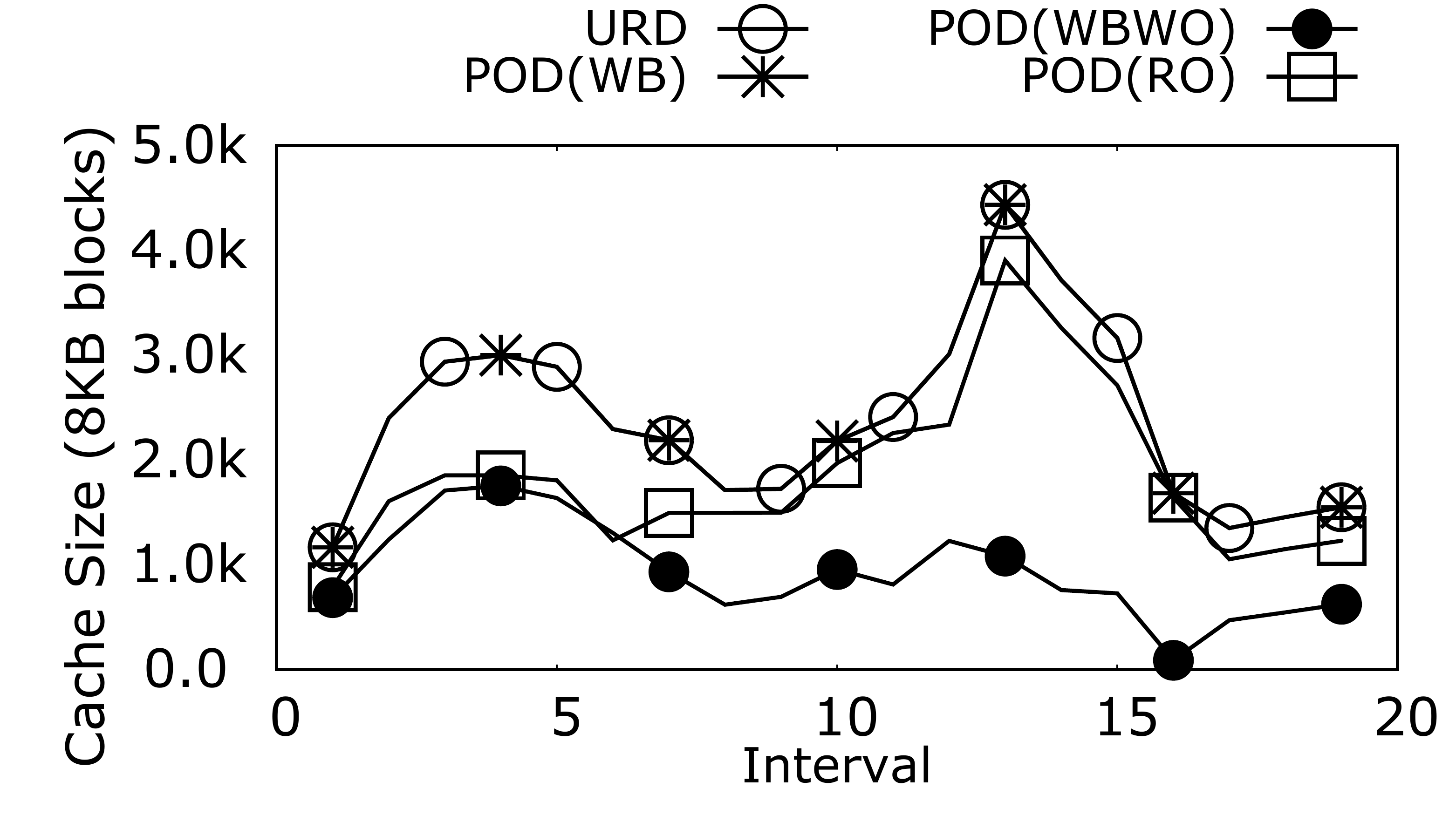}%
		\label{fig:hm1_20}}
	\hfil
	\subfloat[VM1: \mdssefr]{\includegraphics[width=.24\textwidth]{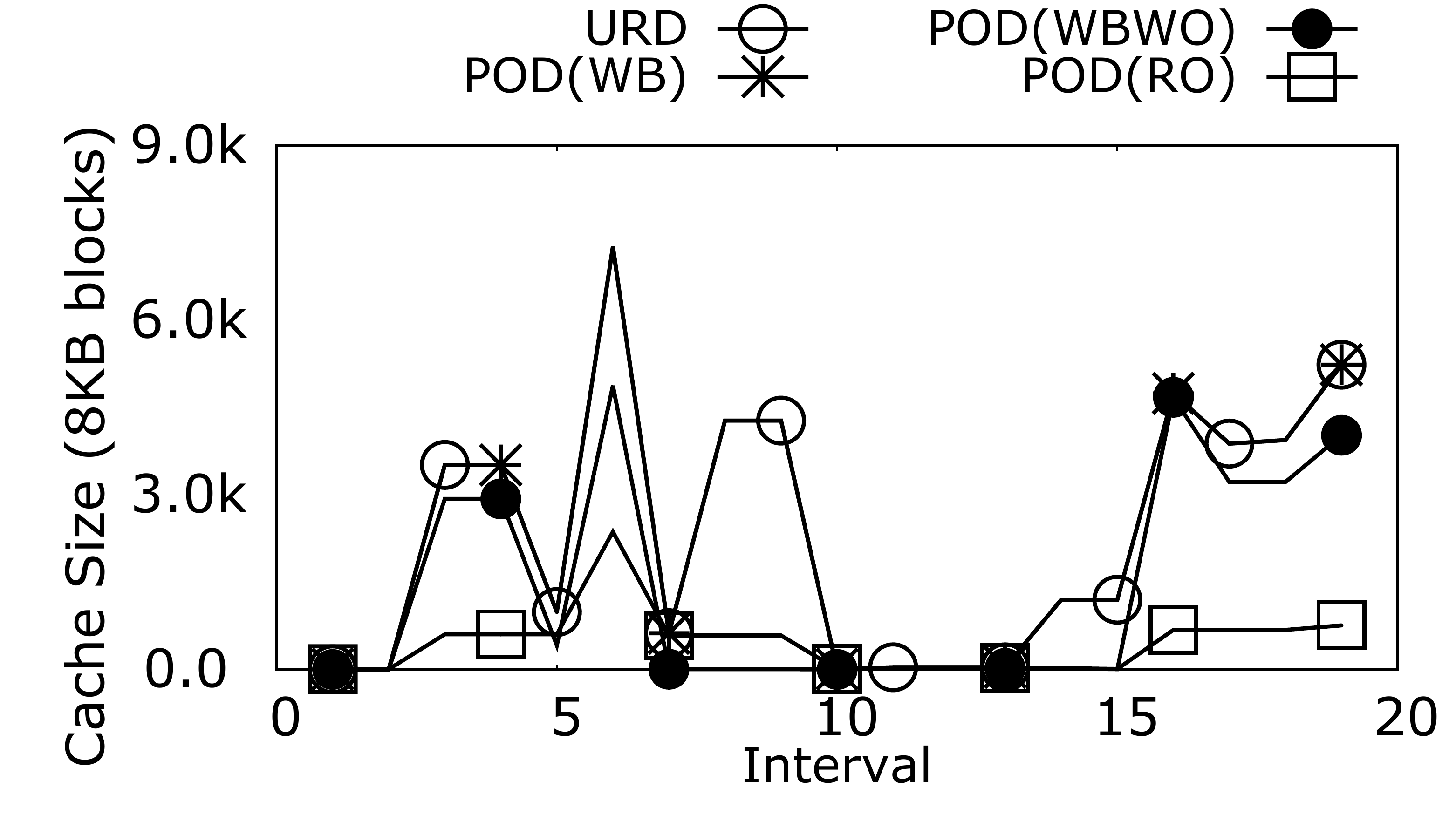}%
		\label{fig:mds0_20}}
	\hfil
	\subfloat[VM2: \proj]{\includegraphics[width=.24\textwidth]{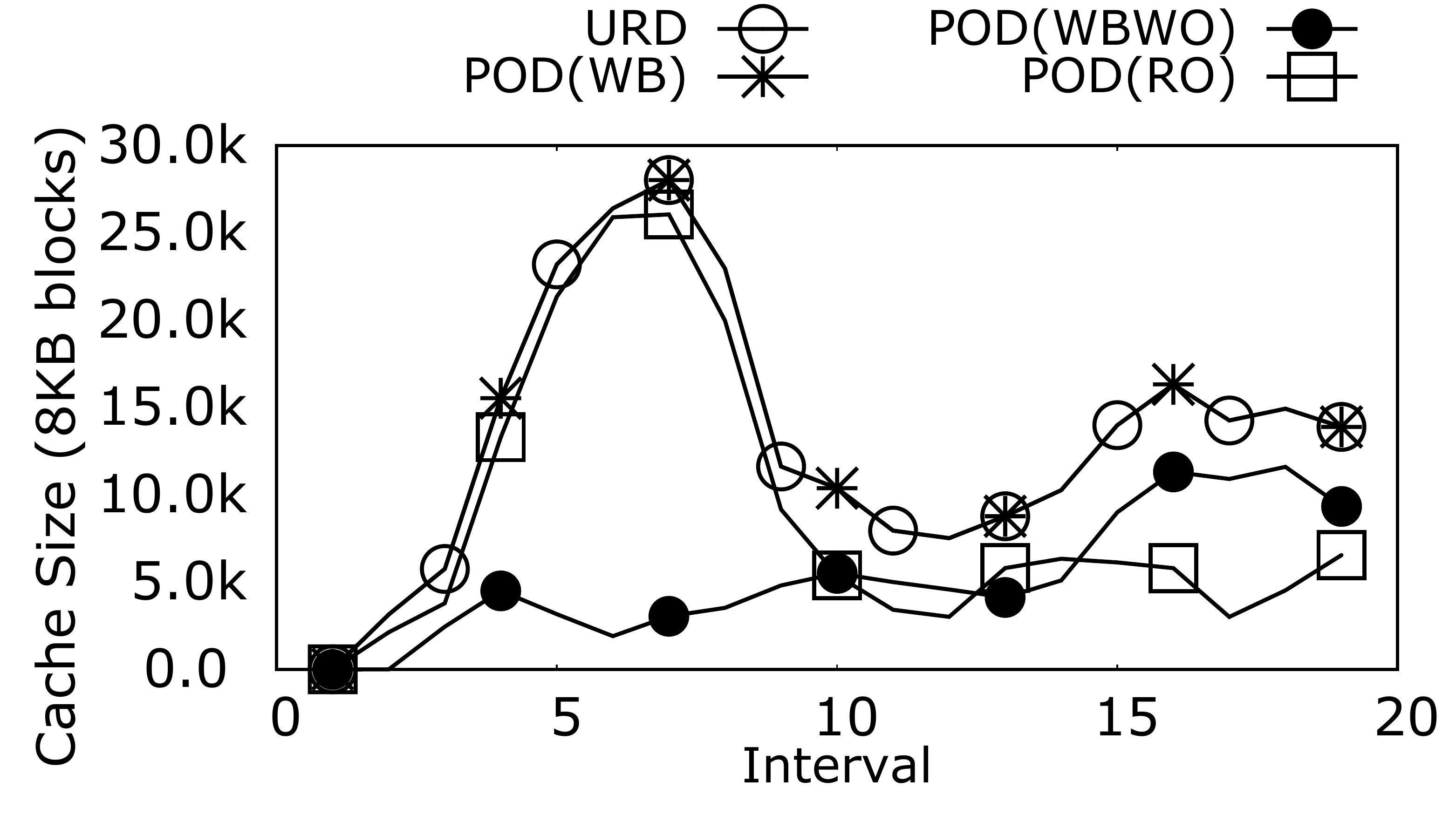}%
		\label{fig:proj0_20}}
	\hfil
	\subfloat[VM3: \rsrchsefr]{\includegraphics[width=.24\textwidth]{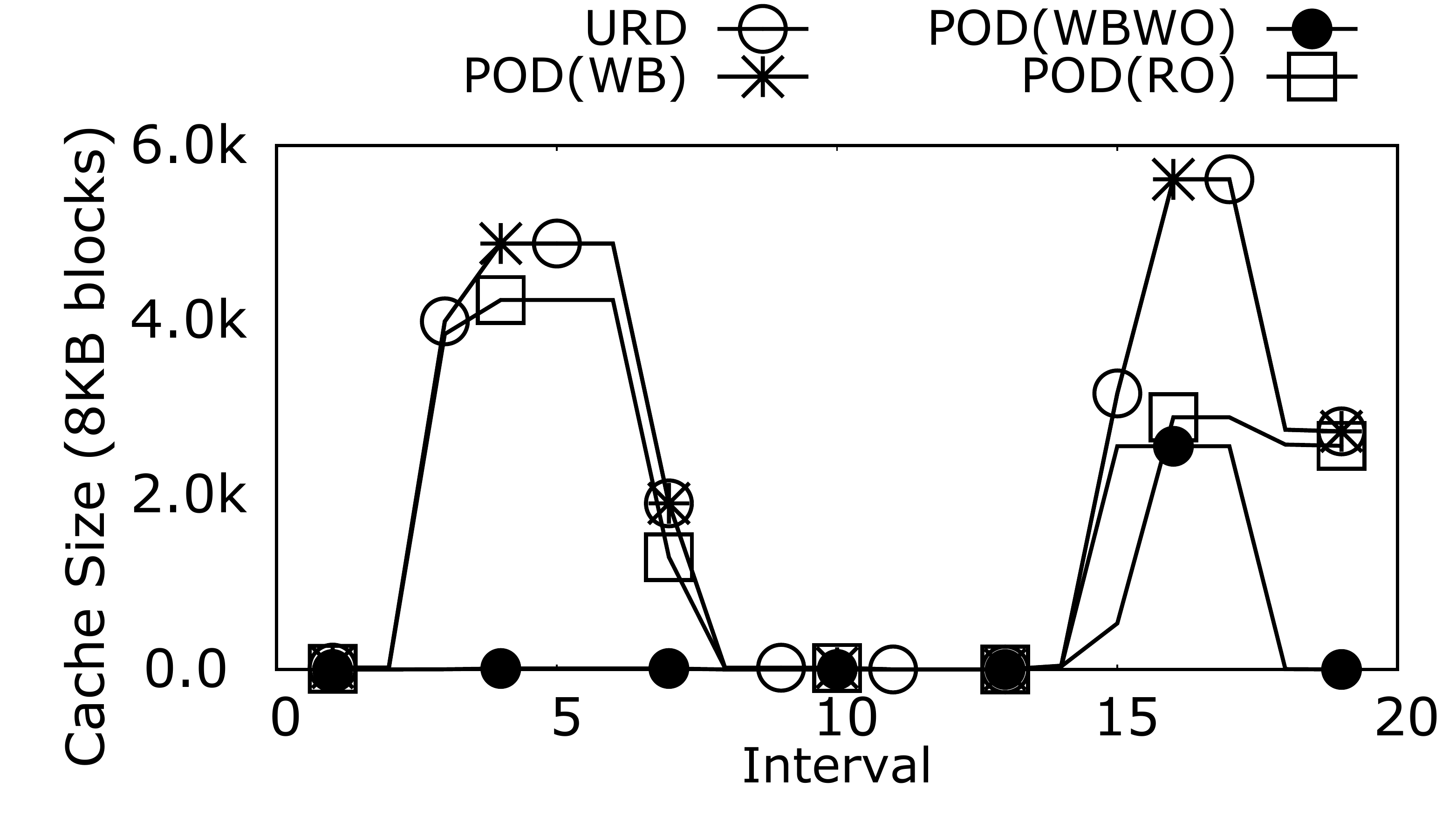}%
		\label{fig:rsrch0_20}}
	\hfil
	\subfloat[VM4: \srcdo]{\includegraphics[width=.24\textwidth]{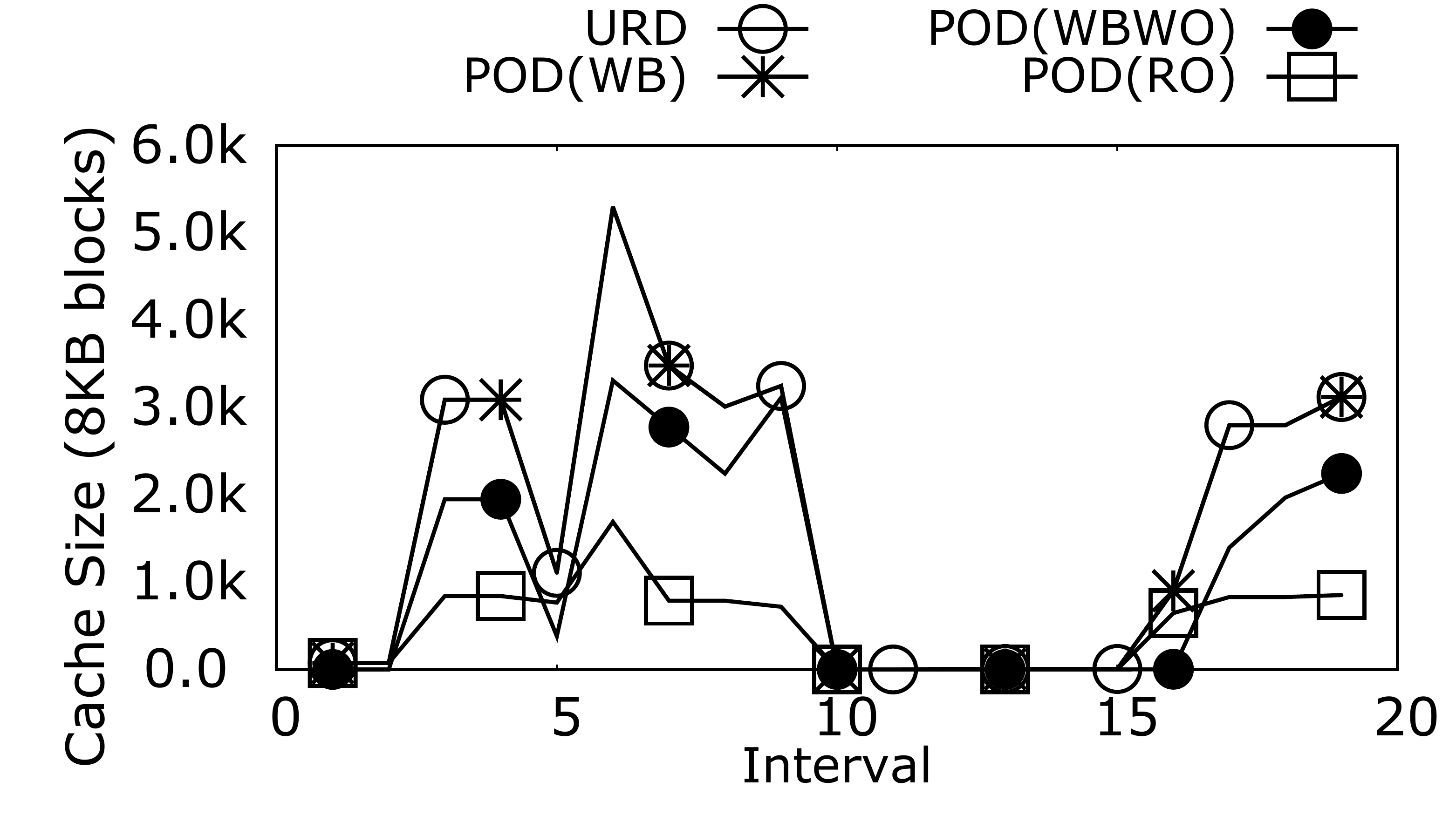}%
		\label{fig:src20_20}}
	\hfil
	\subfloat[VM5: \stg]{\includegraphics[width=.24\textwidth]{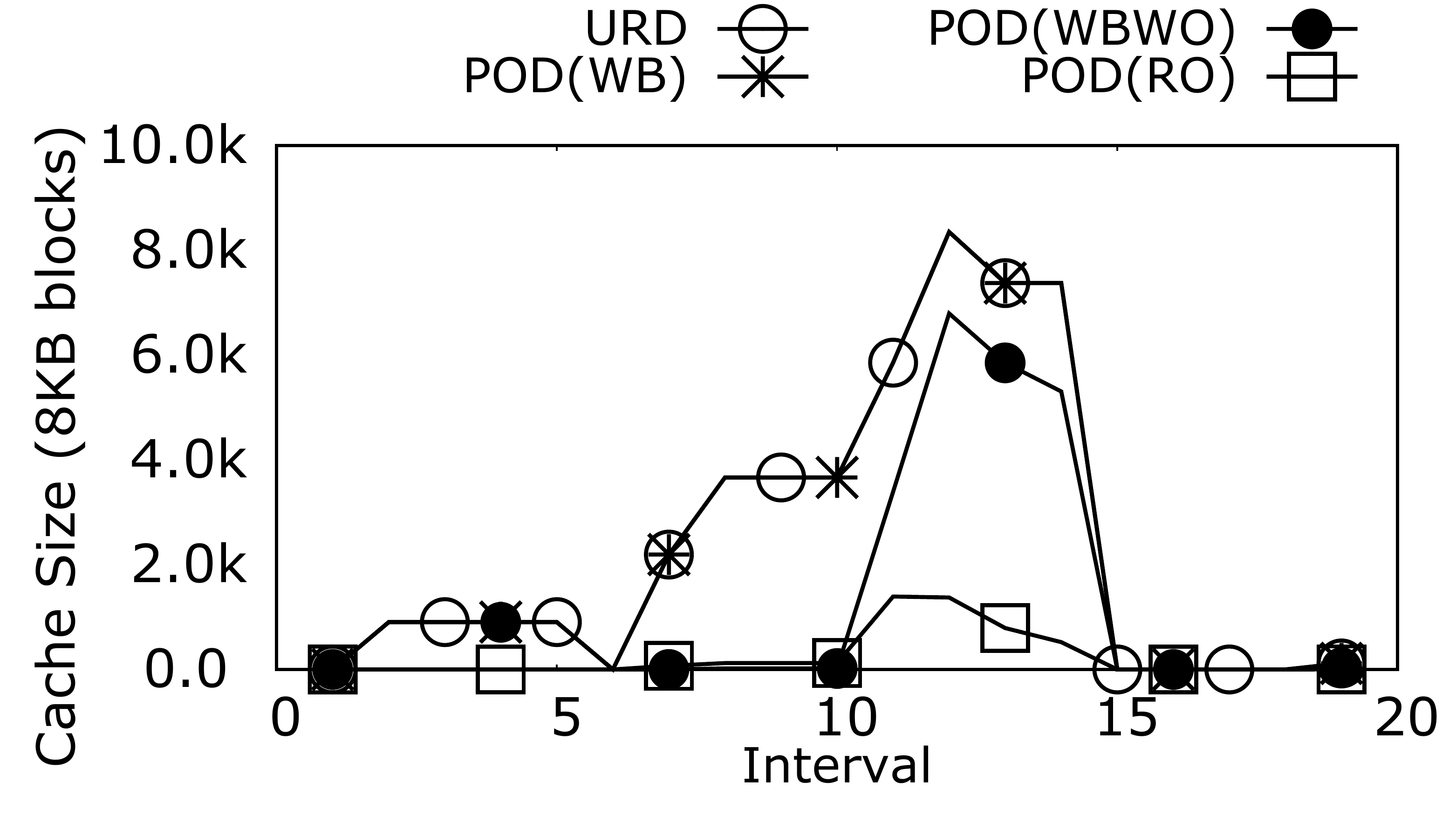}%
		\label{fig:stg1_20}}
	\hfil
	\subfloat[VM6:‌ \usr]{\includegraphics[width=.24\textwidth]{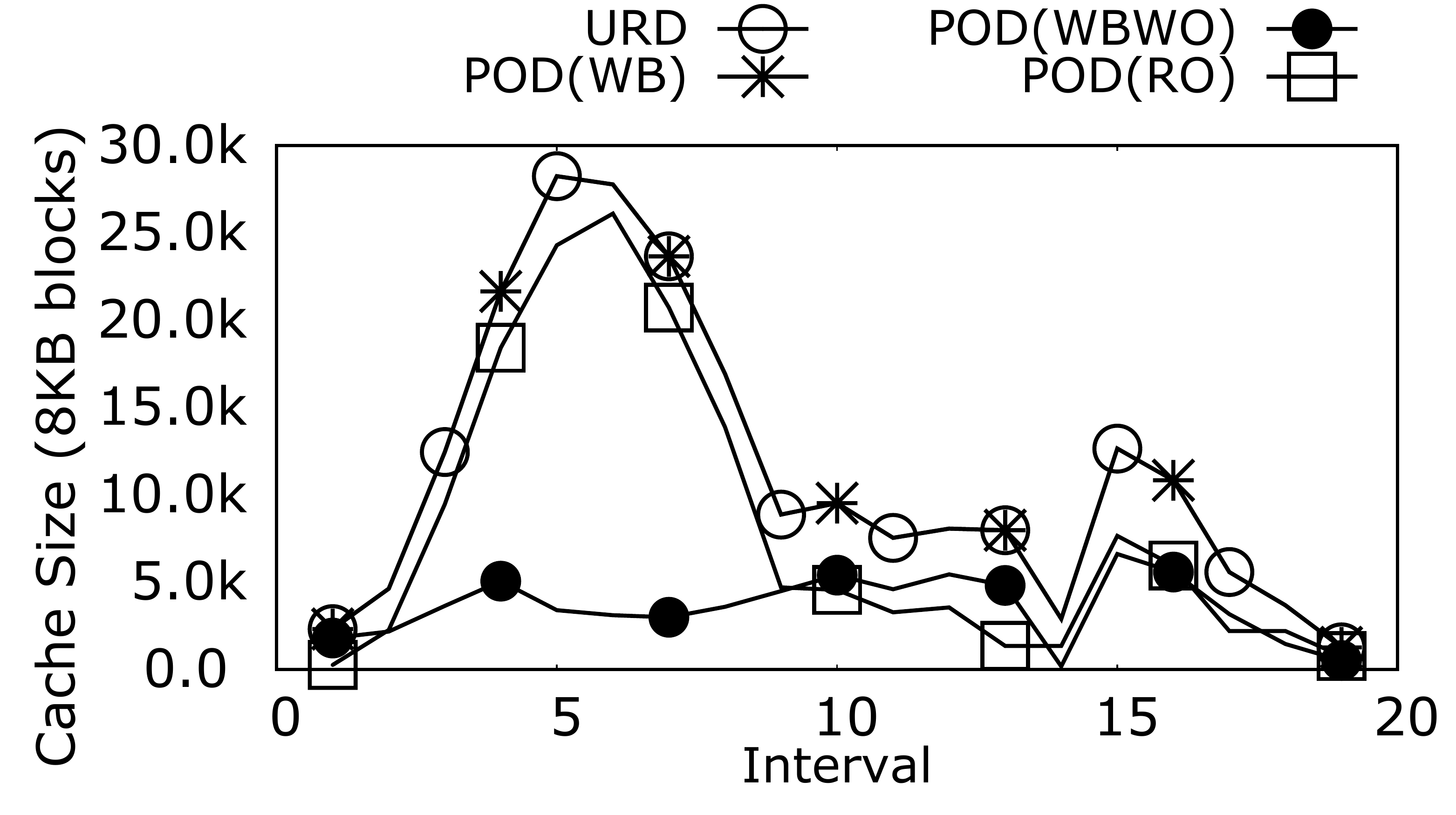}%
		\label{fig:usr0_20}}
	\hfil
	\subfloat[VM7: \mdsyek]{\includegraphics[width=.24\textwidth]{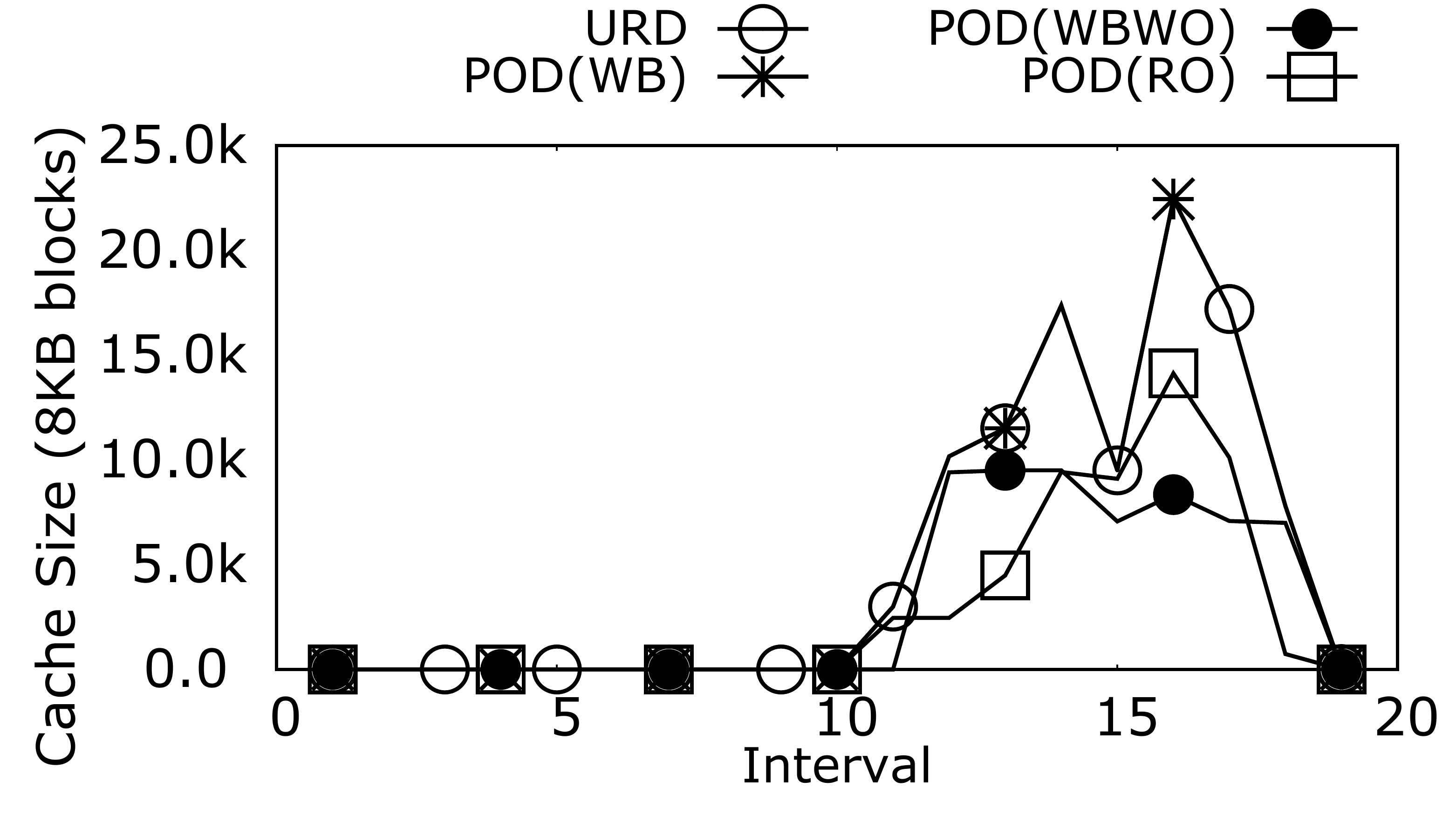}%
		\label{fig:mds1_20}}
	\hfil
	\subfloat[VM8: \srcyek]{\includegraphics[width=.24\textwidth]{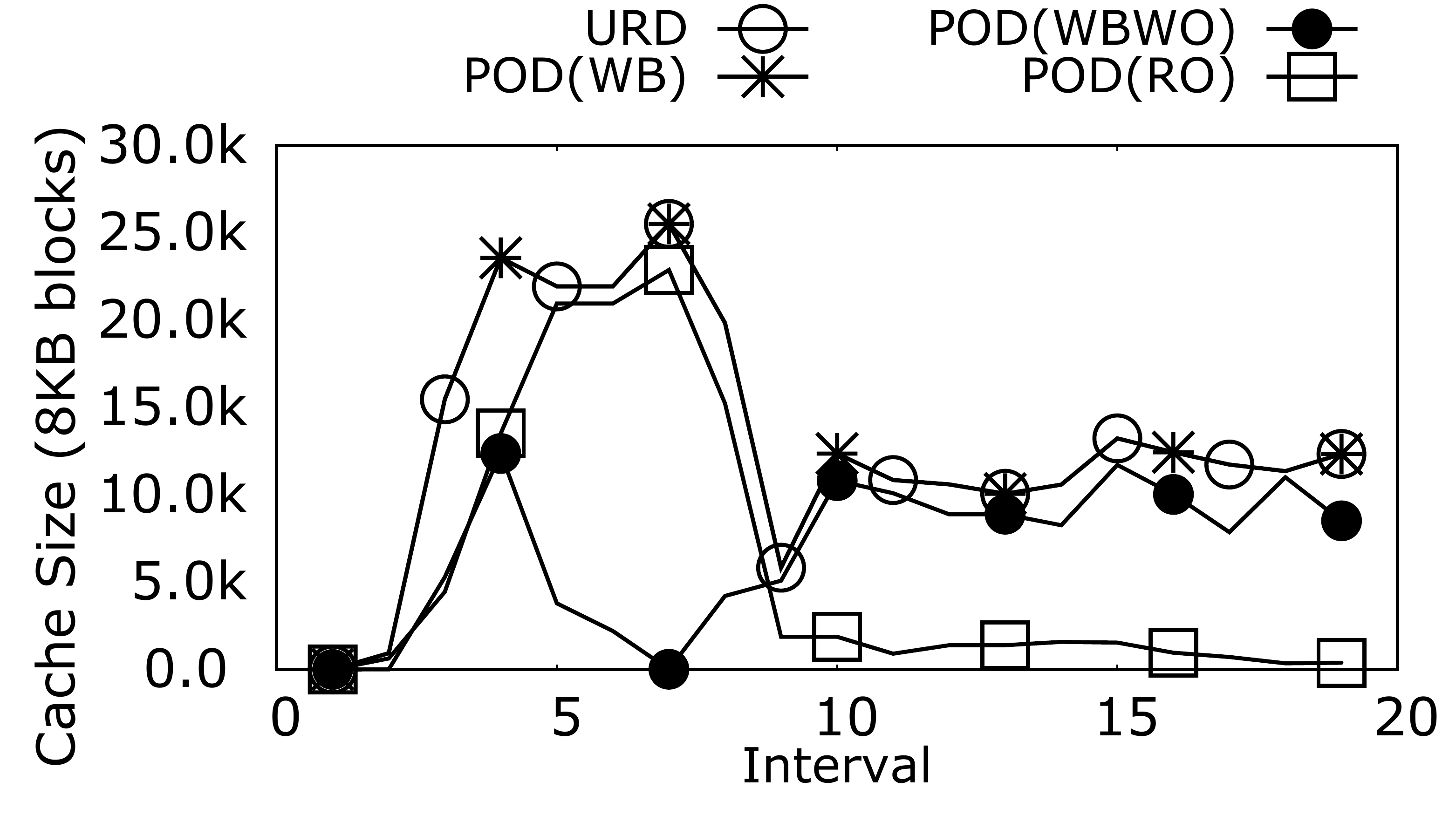}%
		\label{fig:src12_20}}
	\hfil
	\subfloat[VM9: \ts]{\includegraphics[width=.24\textwidth]{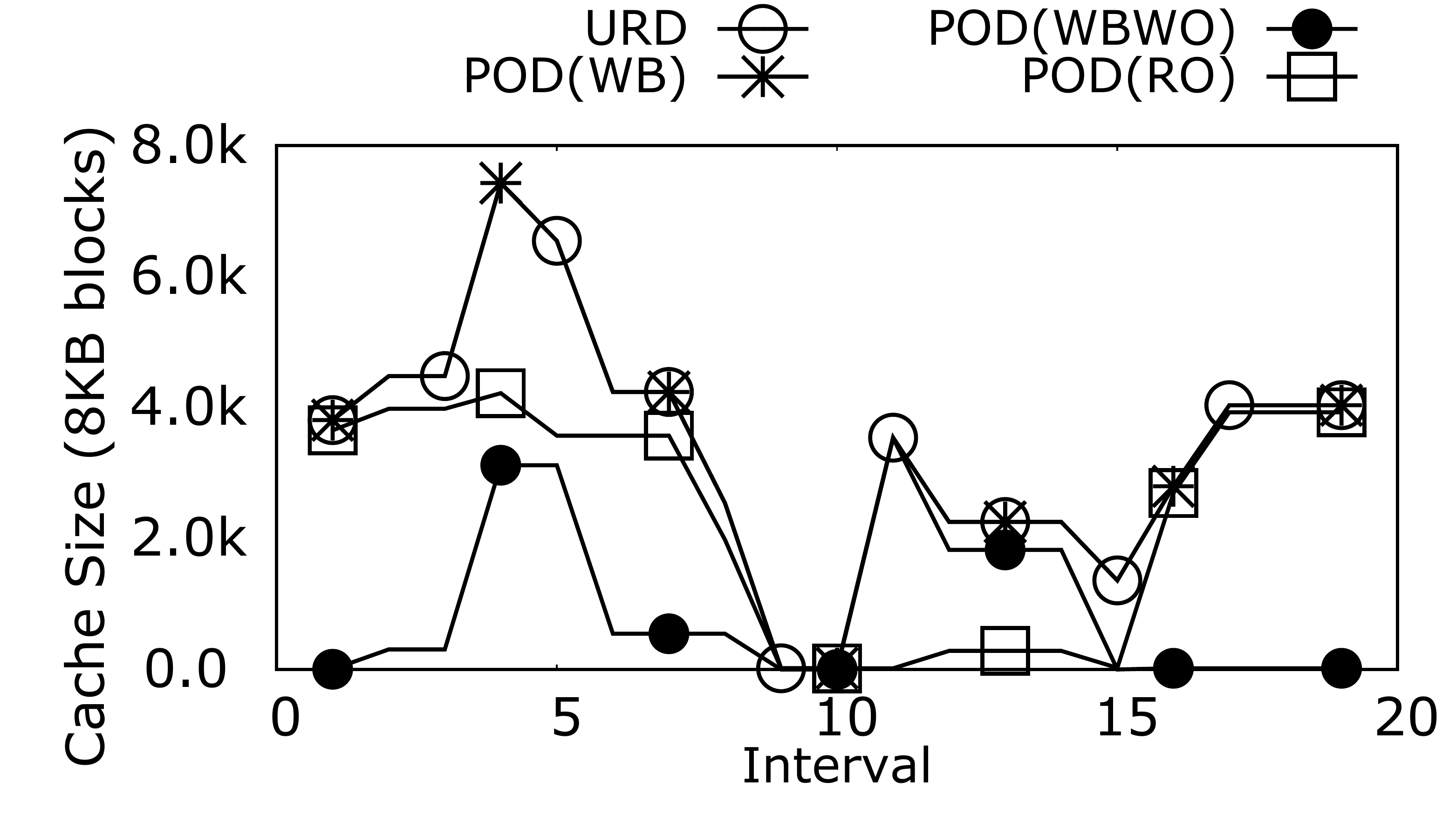}%
		\label{fig:ts0_20}}
	\hfil
	\subfloat[VM10: \wdev]{\includegraphics[width=.24\textwidth]{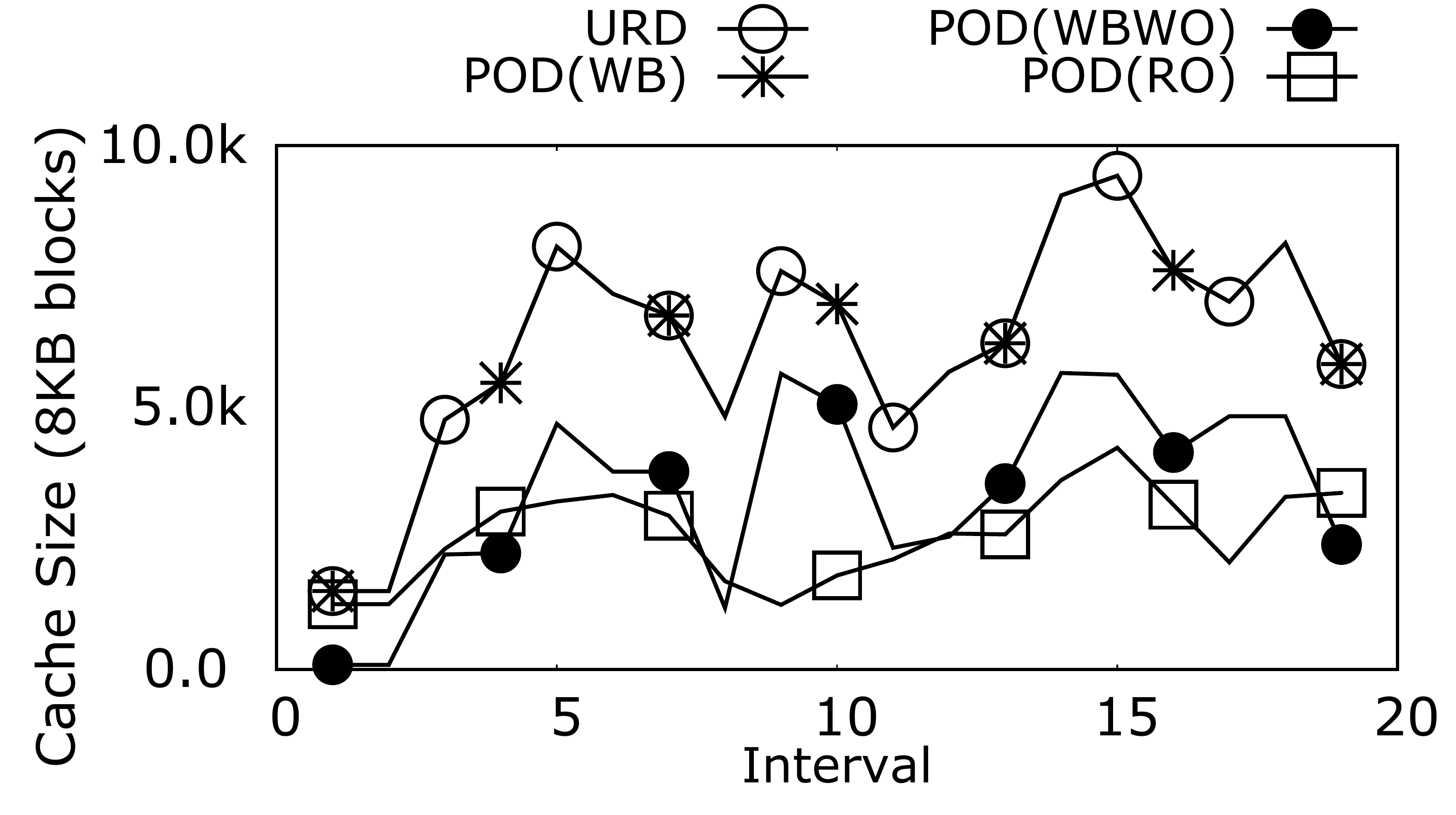}%
		\label{fig:wdev0_20}}
	\hfil
	\subfloat[VM11: \webse]{\includegraphics[width=.24\textwidth]{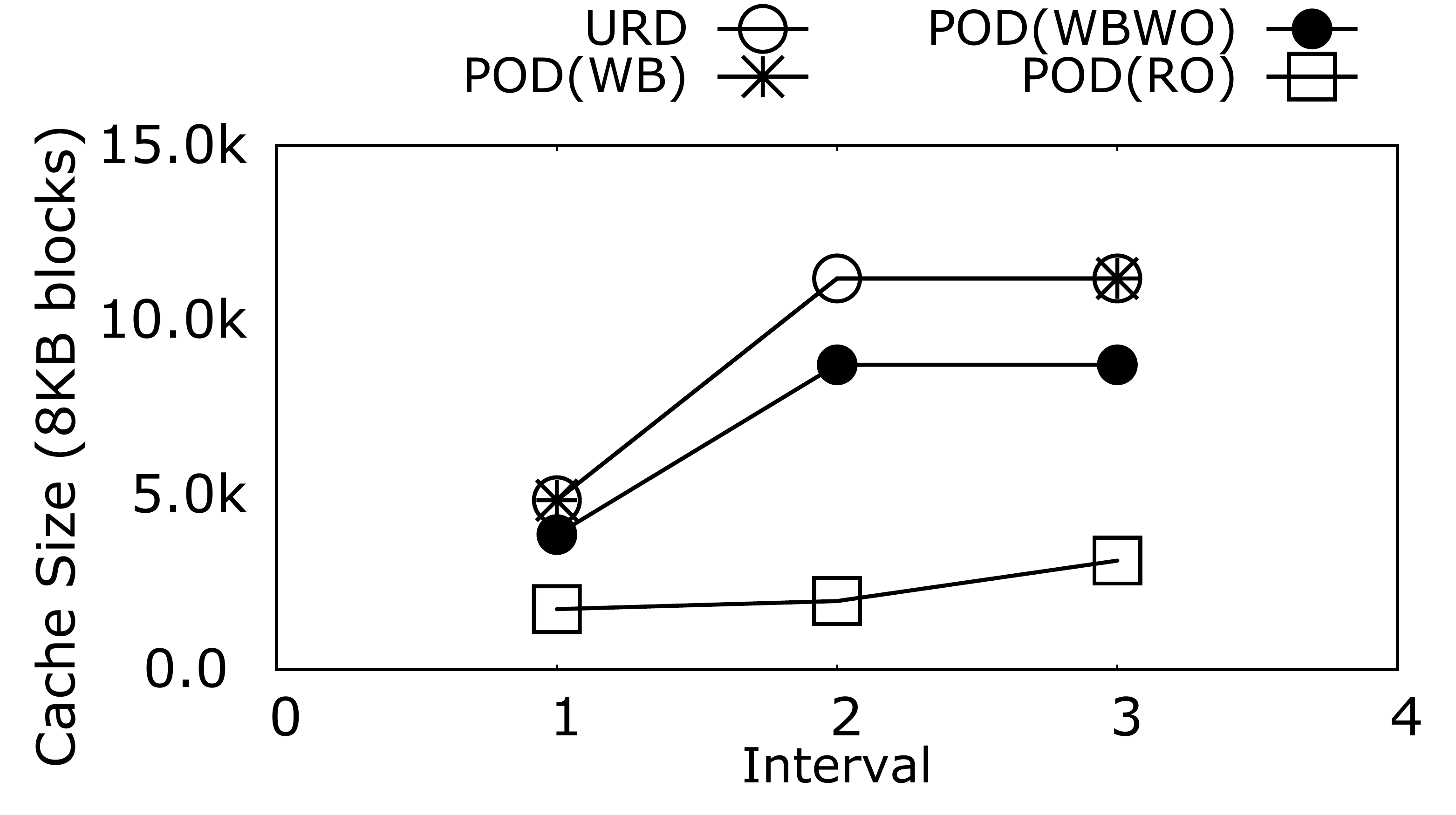}%
		\label{fig:web3_20}}
	
	\caption{Allocated cache sizes by \techname{} and ECI-Cache (URD: ECI-Cache, \URDP{}(RO): \techname{} at the DRAM level, and \URDP{}(WBWO): \techname{} at the SSD‌ level).}
	\label{fig:cost_impv}
\end{figure*}

\subsection{Experimental Setup}
\label{sec:setup}
We conduct comprehensive experiments on a {real} system including a 4U rack mount SuperMicro server with 1) 6x Seagate 4TB SAS 7.2K HDDs in RAID5 (5+1) configuration {at the} disk subsystem level, 2) 4x Samsung 2TB 863a SATA‌ SSDs configured {as} RAID10 (2+2) {at the} SSD cache level, 3) 128GB Samsung DDR4 DRAM where 100GB is used {at the} DRAM cache level, 4) LSI9361i PCIe RAID controller {to apply the}  RAID configuration, 5) 32x Intel(R) Xeon (R) $2.1$GHz CPU cores, and 6) {a} SuperMicro X10i motherboard. We integrated \techname{} with QEMU as {the} open-source hypervisor running on {the} CentOS7 operating system (kernel version: 3.10.327.36.3).\footnote{We disable the buffer cache and set the configuration of the block layer to the default mode with \emph{merge} option enabled with a 128-entry device queue size.} {We implement \URDP{} by modifying the source code of PARDA \cite{parda}. }

We run VMs with {the} \emph{CentOS} and \emph{Ubuntu} operating systems on the hypervisor. {Each} VM is configured with 1) 50GB disk space, 2) 1GB memory, and 3) two virtual CPUs. 
{\techname{} estimates and allocates the I/O cache size for each VM based on the I/O {patterns} of  workloads {running} on the VMs. We run MSR traces from SNIA \cite{msr,narayanan2008write}  in the device layer}, including more than ten real application workloads with different I/O {characteristics,} such as hm\_1 (hardware monitoring), mds\_0 (media server), mds\_1, src2\_0 (source control), src1\_2, stg\_1 (web staging), ts\_0 (terminal server), wdev\_0 (test web server), web\_3 (web/SQL server), rsrch\_0 (research projects), usr\_0 (user home directories), and proj\_0 (project directories).
To show the {effectiveness} of \techname{}, we compare {it} with the latest state-of-the-art I/O caching scheme in virtualized platforms, ECI-Cache \cite{ahmadian2018eci}. To do so, we implemented this scheme in our {real testing} platform and evaluated {it} with the same experiments.
{The evaluation of ECI-Cache and \techname{} are performed via the default configuration of the device layer with enabled \emph{request merge} option for a 128-entry device queue size and disabled buffer cache.\footnote{{Note that ETICA and ECI-Cache has no management on the buffer cache and they only perform in the device layer.}}}



\subsection{Cache Size Improvement}
\label{sec:cost}

In this section, we {empirically} show {that} \techname{} {reduces the allocated cache {space} to {each} VM}. {Our} experiments  show how \techname{} estimates {a} smaller cache size compared to {the} previous state-of-the-art cache space partitioning scheme, ECI-Cache, while preserving the performance (i.e., latency) of the running VMs (Section \ref{sec:perf}).
\techname{} employs the \URDP{} metric in cache size estimation, which effectively allocates smaller cache sizes to the VMs {by taking into account} both \emph{request type} and \emph{cache write policy} {(as we showed in Section \ref{sec:urd+})}. The  state-of-the-art scheme, ECI-Cache, works based on URD; {we have shown that this} metric ignores the impact of cache write policy and only considers the request type in reuse distance estimation and {thus} overestimates the cache size of the VMs with RO and WBWO policies ({see} Section \ref{sec:urd+}). 
We run 12 VMs with different types of workloads on the hypervisor and {measure} the cache size estimation {of} \techname{} and ECI-Cache. To do so, we set different cache write policies to the VMs: 1) WB, 2) WBWO, and 3) RO. We estimate the cache size of the VMs in predefined intervals (after observing $10,000$ I/O requests) by \URDP{} and URD which are used by \techname{} and ECI-Cache, respectively.

\begin{figure}[!t]
	\centering
	\hspace*{-.5cm}
	\includegraphics[scale=0.32]{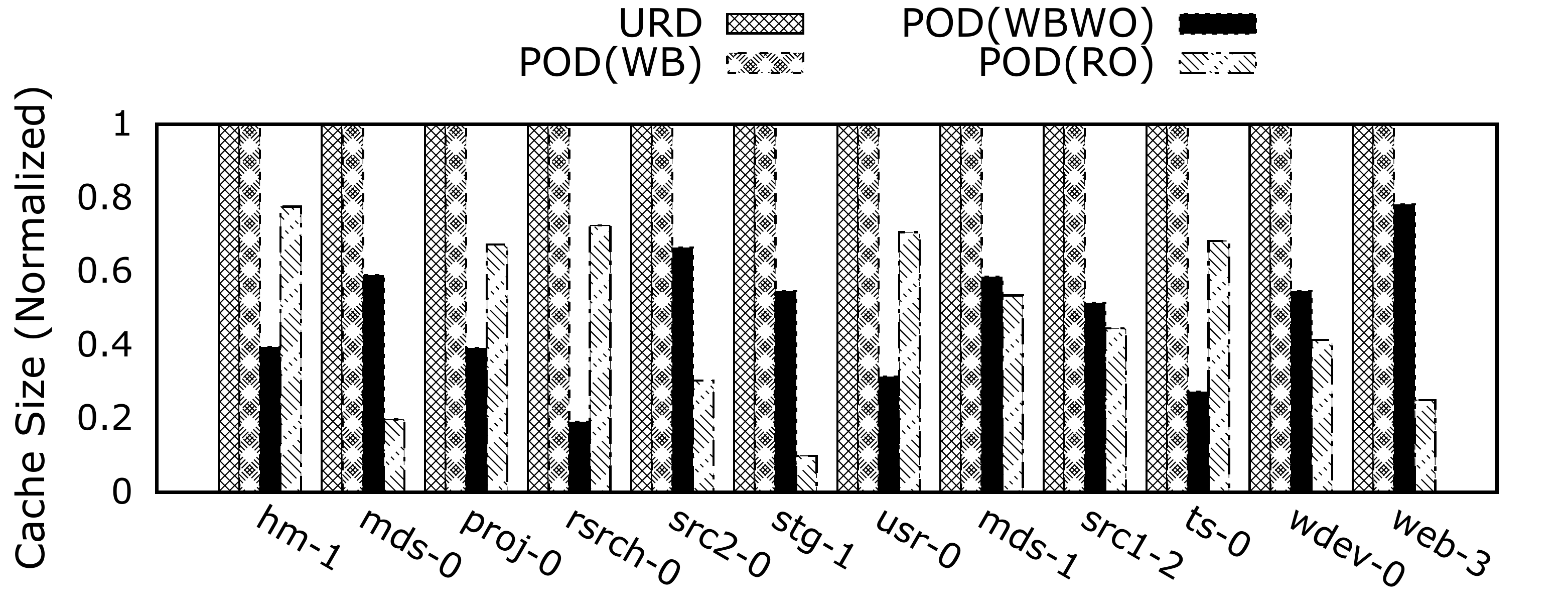}
	\caption{Average cache size allocated by ECI-Cache and \techname{} (URD: ECI-Cache, \URDP{}(RO): \techname{} in DRAM level, and \URDP{}(WBWO): \techname{} in SSD‌ level).}
	\label{fig:avg_size}
\end{figure}

Fig. \ref{fig:cost_impv}  shows {the VM cache size estimations of} \URDP{} and URD {for} different cache write policies. Unlike \URDP, URD {works independently of the cache write policy: that is, URD} works {\emph{exactly}} the same {for} the caches with WB, WBWO, and RO policies,  and hence, we {show} {the} URD results {with} one single line.
To {make} the {differences} of cache sizes estimated by the proposed and previous schemes more visible {in the figures}, we  show {only} the first 20 intervals of the workloads. 
\footnote{{We run the experiments for more than 500 intervals (until the workload finishes) and observe very similar behavior across the entire execution.}}
In addition, Fig. \ref{fig:avg_size} compares the {\emph{average cache sizes}} allocated by ECI-Cache (URD) and \techname{} (\URDP{}) for the VMs {over their entire runs}.
We make {five} main observations:

\begin{enumerate}[leftmargin=*]
	\item 
	The URD metric allocates the {\emph{same}} size for the caches with different write policies (i.e., RO, WBWO, and WB). This is because URD {\emph{ignores}} the cache write policy and \emph{only} considers  request type in reuse distance calculation. {As we have clearly shown, this} scheme {allocates} cache blocks for requests {that} would {\emph{not}} be buffered in cache. For instance, {a} WBWO cache does not buffer read {requests}{, but} URD {reserves} cache blocks for such {requests} {in its reuse distance calculation}. Similarly, URD {allocates} cache blocks of {an} RO cache for write  {requests} that {bypass the} {RO} cache.
	\item 
	The {amount of}  cache space {allocated} by \URDP{} in both RO and WBWO policies is considerably {smaller} {(i.e., by 51.7\%, on average)}  {than that allocated by URD}. This is due to the fact that in {an} RO cache, \URDP{}  considers {only} RAR accesses in {the} reuse distance calculation and does {\emph{not}} {allocate} cache blocks for writes. {In} a WBWO cache{, \URDP{} considers} only {the} reuse distance of RAW accesses and {allocates no cache block} for cold reads (i.e., read {misses}). {In contrast}, URD considers both RAR and RAW accesses {\emph{without}} taking into account the {effect} of cache write policy {on what gets cached and what does not}. Unlike \URDP{}, the URD metric allocates cache blocks for {both} writes and reads, which would {\emph{not}} be buffered in RO and WBWO caches, respectively.
	\item 
	In the intervals where RAR accesses are involved in cache size estimation (e.g., intervals 10 to 15 in \hm{}, 3 to 10 in \proj{}, and 3 to 6 in \rsrchsefr{}), URD and \URDP{} for {the} RO cache (\URDP{}(RO)) have the same behavior but \URDP{} estimates smaller cache sizes ({because it does} {\emph{not} allocate} cache blocks for write requests). Similarly, in the intervals {where}  RAW accesses are involved (e.g., intervals 1 to 3 in \webse{}, 0 to 8 in \ts{}, and all intervals of \wdev{}), URD and \URDP{} for {the} WBWO cache (\URDP{}(WBWO)) have the same pattern {but} \URDP{} {estimates a much smaller cache size because it does}  {\emph{not}} {allocate} blocks for cold reads in {the} WBWO cache.
	\item
	In the cache with {the} WB policy, \URDP{} and URD estimate the same cache {size}. This is because WB cache buffers {\emph{both}} read and write requests and hence, both schemes {allocate} cache blocks for both types of requests.
	\item 
	{The} {\URDP{} calculation imposes up to 0.83\% performance overhead{, which is similar to the overhead imposed by} ECI-Cache due to calculating URD {(not shown in the figures)}. We observe that both \URDP{} and URD {calculations} have the same performance overhead on the running VMs.}
\end{enumerate}

We conclude that {\URDP{} provides better, more efficient size estimation for caches that employ RO and WBWO policies, thereby reducing waste of space in an I/O cache}.

\subsection{I/O Latency Improvement}
\label{sec:perf}
{In this section, we {empirically} show {that} \techname{} improves the overall performance (in terms of I/O latency) of the running VMs.}
In the experiments of this section, we run 12 VMs with different workloads on the hypervisor integrated with \techname{}  and examine the performance {in terms of I/O latency} of the running VMs.
{We evaluate the system performance using \techname{} {under} two conditions: 1) without applying {the} promotion/eviction scheme (denoted as \techname{}-NPE) and 2) with applying our proposed promotion/eviction scheme (denoted as \techname{}-Full).}
We perform the same experiments on {the} ECI-Cache architecture {for performance comparison}.
{We compare the performance of these two schemes by assuming similar total cache spaces, where the {total} space of SSD+DRAM in \techname{} is {\emph{equal}} to the space of SSD in ECI-Cache.}
We recalculate the cache sizes after observing $10,000$ I/O requests  and we {also} promote/evict data blocks  after observing $1,000$ I/O requests {(in Section \ref{sec:analysis} we evaluate the impact of promotion/eviction intervals on both the performance and endurance improvement {of} \techname{})}.  

\begin{figure}[!h]
	\centering
	\hspace*{-.5cm}
		\includegraphics[scale=0.31]{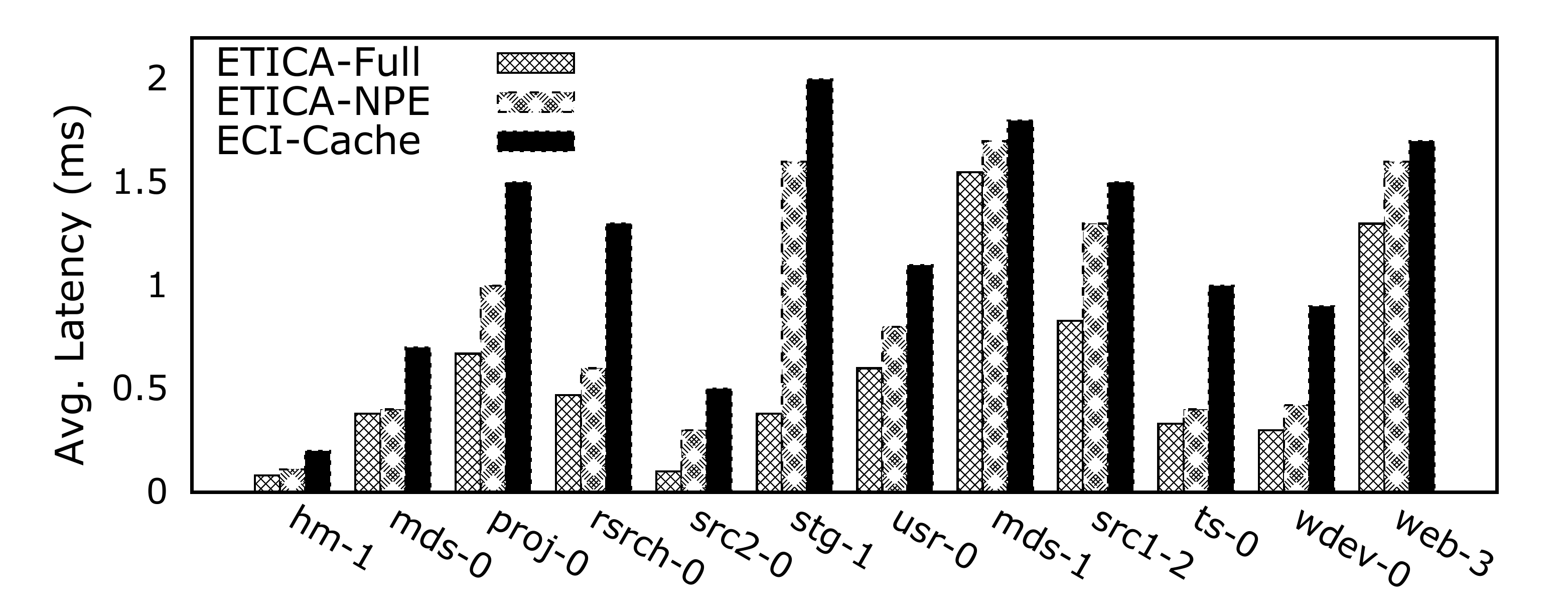}
	\caption{Average latency of VMs with \techname{} and ECI-Cache (\techname{}-NPE: without promotion/eviction and \techname{}-Full: with promotion/eviction).}
	\label{fig:avg_latency}
\end{figure}

\begin{figure}[!h]
	\centering
	\includegraphics[scale=.31]{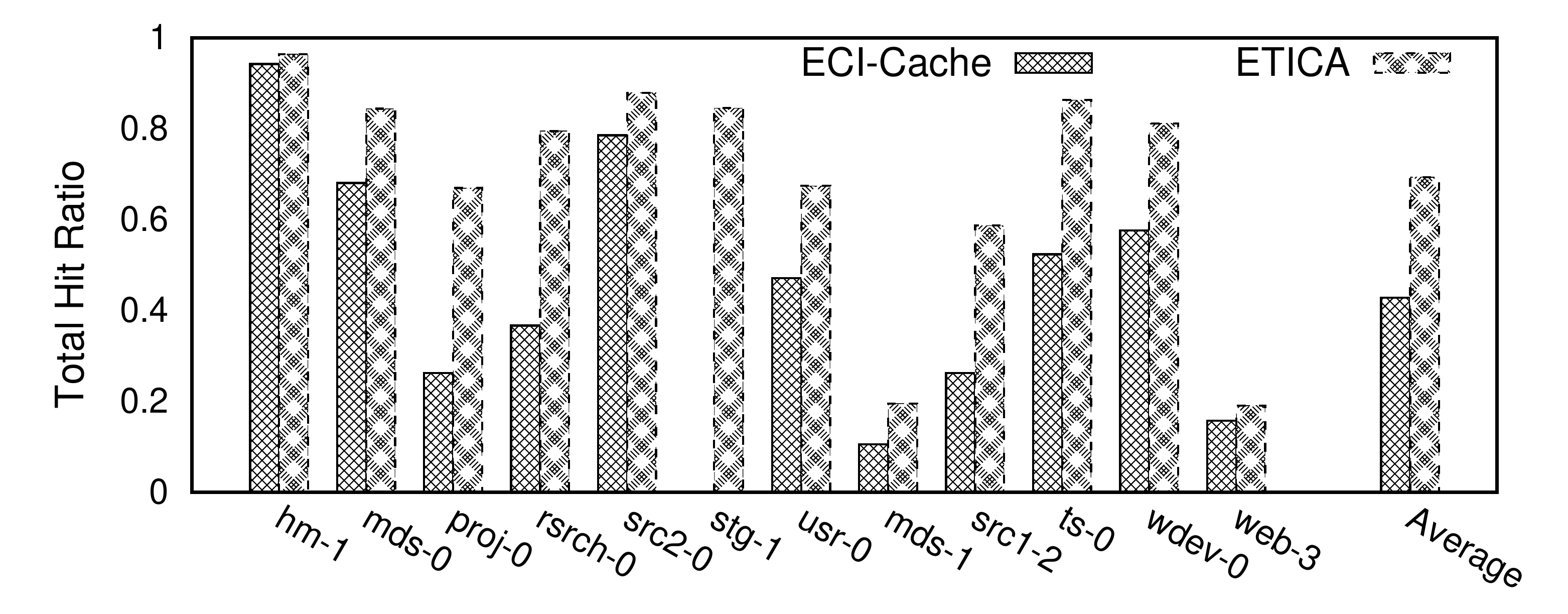}
	\caption{Total hit ratio of VMs with \techname{} and ECI-Cache.}
	\label{fig:hit_ratio}
\end{figure}

{Fig. \ref{fig:avg_latency} and Fig. \ref{fig:hit_ratio} show the results of the experiments, comparing the average latency and total hit ratio of the running VMs with \techname{} and ECI-Cache.}
We make five  observations:
\begin{enumerate}[leftmargin=*]
	\item \techname{}, compared to ECI-Cache, improves performance by up to 64\%  (45\%, on average). This is mainly due to the improvement in the latency of read requests. {{Even} without applying {our} proposed promotion/eviction scheme, \techname{} improves the I/O performance by 27\%, on average.}
	
	\item {Performance enhancement by \techname{} is due to 1) hit ratio improvement and 2) DRAM high I/O performance. As shown in Fig. \ref{fig:hit_ratio}, \techname{} provides 30\% higher hit ratio compared to ECI-Cache.  In this case, allocating DRAM to ECI-Cache cannot close the performance gap between \techname{} and ECI-Cache.}
	
	\item In VM3 running the \mdsyek{} workload, {compared to ECI-Cache,}{ \techname{}-NPE and \techname{}-Full achieve {only} 1\% and 13\% performance improvement, respectively}. This is due to the sequential read access pattern of the running workload with low locality of reference. As we observe in the experiments of Section \ref{sec:cost}, both ECI-Cache and \techname{} assign almost zero cache space for this workload in the first 10 intervals. Using a cache for such workloads with sequential (streaming) read accesses {with low locality} {is unlikely to} provide {significant} performance improvement since {most} requests would be supplied by {the} disk subsystem.
	
	\item The \usr{} workload mainly consists of write requests, which are supplied by the SSD cache in both \techname{} and ECI-Cache schemes. \techname{}-Full, compared to ECI-Cache, improves the I/O performance of this workload by 45\%  since it buffers RAW requests at the DRAM level and does not evict popular written data blocks from the SSD {(\techname{}-NPE improves the performance of this workload by 27\%).}
	
	\item The running workload in VM1, \hm{}, mainly includes random read accesses with high locality of reference. \techname{} obtains a high hit rate from DRAM (about 91.6\%) for this workload and improves  I/O performance by 42\%. ECI-Cache achieves a similar hit rate in the SSD, but the higher latency of the SSD compared to DRAM leads to lower performance by ECI-Cache.
	
	\item \techname{} achieves the highest performance improvement for the \srcdo{}, \ts{}, and \wdev{} workloads (53\%, 58\%, and 64\%, respectively, over ECI-Cache). These workloads include a small number of write operations with a large number of RAW (and also RARAW) re-references, which leads to promoting a small number of written data blocks from the SSD to the high-performance DRAM and {serving} further accesses to them {with} much lower latency than ECI-Cache. 
\end{enumerate}

We conclude that \techname{} provides 45\% higher I/O performance than ECI-Cache. This improvement is mainly due to 1) the performance improvement of read requests (especially the requests that are supplied by {the} DRAM level) and 2) not evicting popular data blocks from {the} SSD level.

\subsection{SSD Endurance Improvement}
\label{sec:endurance}
\begin{figure}[!h]
	\centering
	\hspace*{-3em}
	\includegraphics[scale=0.32]{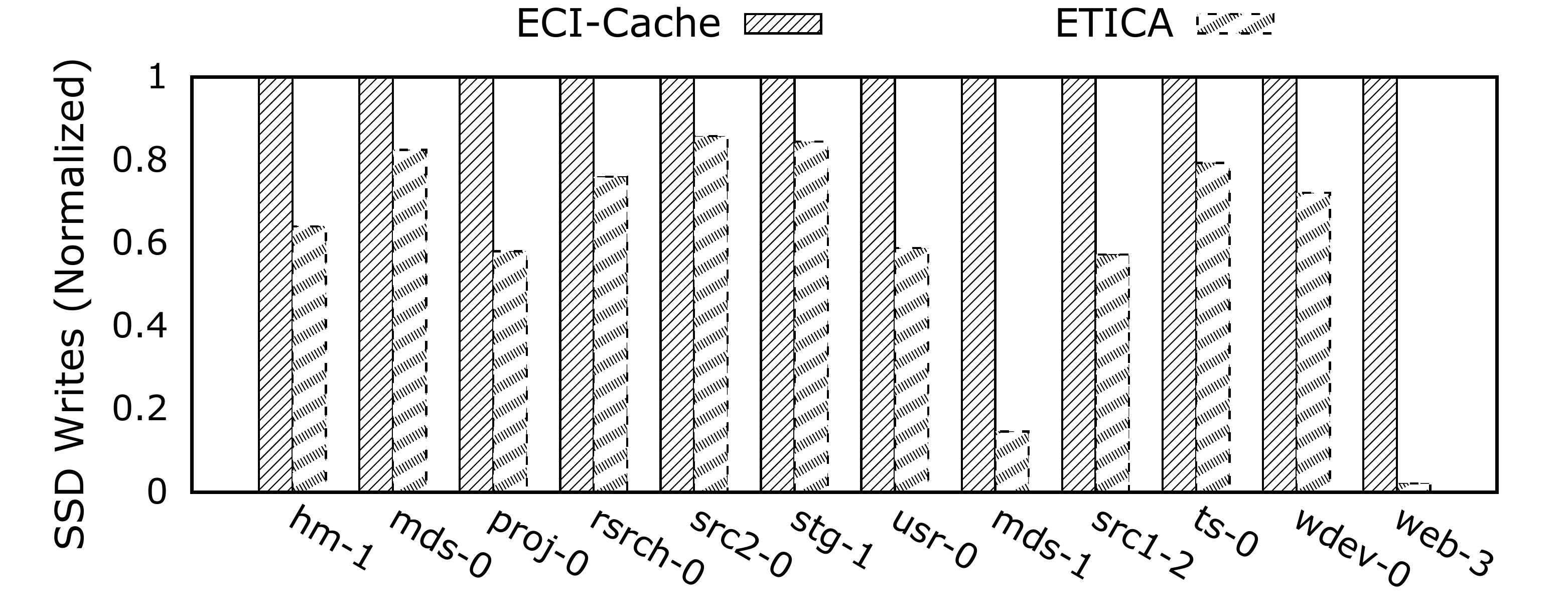}
	\caption{Number of write operations performed on the SSD cache by ECI-Cache and \techname{}.}
	\label{fig:endurance}
\end{figure}

In this section, we evaluate {the} SSD endurance  improvement of \techname{} over ECI-Cache. We measure  endurance \emph{{with the} number of writes into the SSD cache} {as our metric} (as discussed {and used} by prior works \cite{anlj17,elasticQ,huang2016improving,liu2014plc,ahmadian2018eci}).
%
Fig. \ref{fig:endurance} compares the endurance of SSD using \techname{} and ECI-Cache (in terms of {the} number of  writes {performed} into the SSD cache). A lower number of writes indicates higher endurance.
We make two major observations:
\begin{enumerate}[leftmargin=*]
	\item \techname{} reduces the number of  writes into the SSD by 33.8\% in average. The maximum endurance improvement {(about 95\%)} is achieved in the \webse{} workload, mainly due to not buffering read misses at the SSD level.‌ This workload  consists {mainly} of read accesses with a small number of write operations (i.e., it is read-intensive). Thus, applying the RO policy at the SSD level, as ECI-Cache does, does \emph{not} help endurance  since {\emph{all}} read misses (i.e., cold reads) would be buffered in the cache, imposing {a} large number of writes into the SSD.
	
	\item \stg{}, \srcdo{}, and \rsrchsefr{} are write-intensive workloads where both ECI-Cache and \techname{} buffer write requests in the SSD cache. \techname{} achieves about 24\%, 14\%, and 16\% endurance improvement over ECI-Cache for these workloads, respectively. This is because \techname{} does {\emph{not}} update the SSD cache on each miss and hence, it reduces the number of cache updates by only promoting popular blocks into the {SSD} cache.
\end{enumerate}

We conclude that \techname{}, compared to ECI-Cache, significantly reduces the number of write operations on  the SSD cache (by 33.8\%, on average) and thus results in improved {SSD} endurance and lifetime. Read-intensive workloads impose a large number of writes into the SSD cache and \techname{} avoids updating the SSD cache with such read requests.
To obtain high performance, \techname{} buffers read requests at the DRAM level instead of at the SSD level. 
Thus, the two-level I/O caching architecture of \techname{} improves both performance and endurance.

\subsection{Performance Analysis When Enabling New VMs}
\label{sec:analysis-32-VM}
	In this section, we show how \techname{} reallocates cache resources for running VMs and  how it affects the performance (in terms of cache hit ratio) when a varied number of VMs is enabled. 
	In these experiments, we set the total available cache space equal to limited number of cache blocks (i.e., $50,000$ cache blocks). In this case, when the total demand of running VMs exceeds the available cache space, \techname{} reduces the allocated cache space with the minimum performance overhead.
	First we start the experiments with running one VM and then at the specific time instances (i.e., intervals 2, 6, and 10) we extend the number of running VMs to 4, 16, and 32. During intervals 0 to 2, we run hm\_1 workload on VM0.\footnote{The running workloads are from SNIA and the detailed information about workload characteristics are available in \cite{msr,narayanan2008write}.} Then at interval 2, VM1, VM2, and VM3 start to run proj\_0, stg\_1, and usr\_0 workloads, respectively. At interval 6 we run the following workloads in VMs 4 to 15: ts\_0, wdev\_0, web\_3, usr\_0,	mds\_0,	usr\_0,	ts\_0,	wdev\_0,	ts\_0,	hm\_1,	src2\_0, and	ts\_0. At the last interval, VMs 16 to 31 start running workloads: ts\_0,	wdev\_0,	web\_3,	usr\_0,	mds\_0,	usr\_0,	ts\_0,	wdev\_0,	ts\_0,	hm\_1,	src2\_0,	ts\_0,	ts\_0,	wdev\_0,	web\_3, and	usr\_0. The hardware and software configurations of the running VMs in these experiments are the same as reported configurations in Section 5.1.

\begin{figure}[!h]
	\centering
	\includegraphics[scale=.34]{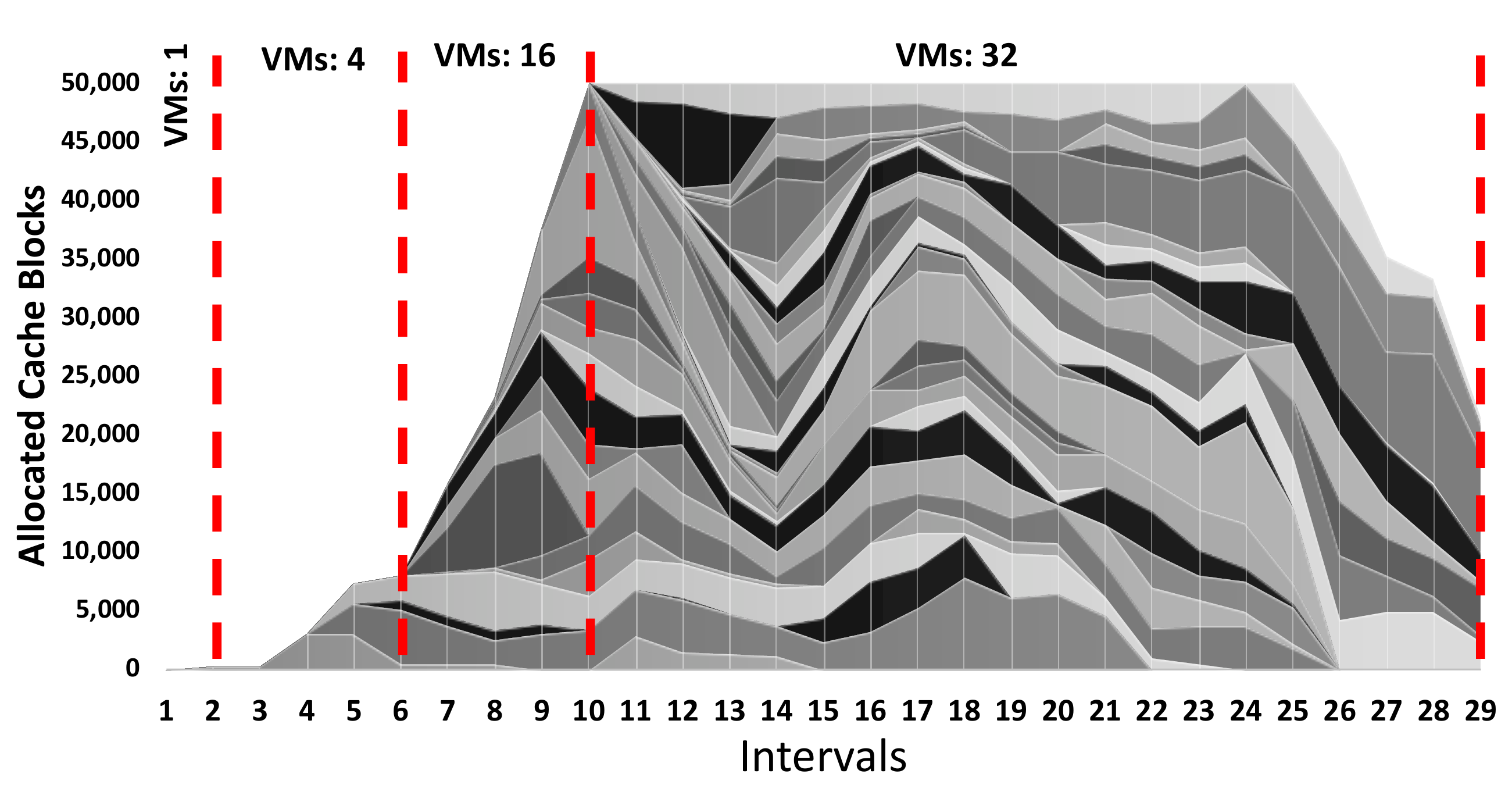}
	\caption{Cache reallocation by \techname{} while enabling different number of VMs (up to 32 VMs, Interval size: 10min).}
	\label{fig:stack-diagram-size1}
\end{figure}
\begin{figure*}[t]
	\centering
	\includegraphics[scale=.65]{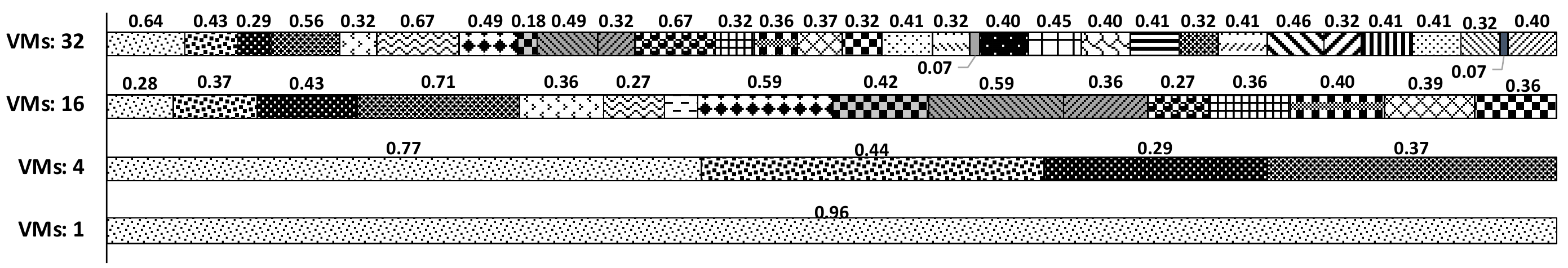}
	\caption{Average performance of VMs (in terms of cache hit ratio) while enabling different number of VMs (up to 32 VMs).}
	\label{fig:Hit-Ratio-32-VM1}
\end{figure*}

	Fig. \ref{fig:stack-diagram-size1} and Fig. \ref{fig:Hit-Ratio-32-VM1} show the results of the experiments. In Fig. \ref{fig:stack-diagram-size1}, we present how \techname{} recalculates new cache sizes for the running VMs based on their demand and available total cache size (equal to $50,000$ cache blocks in the experiments). Fig. \ref{fig:Hit-Ratio-32-VM1} shows the average cache hit ratio of the running VMs while we run 1, 4, 16, and 32 online VMs. 

	It can be seen that in the intervals 1 to 10, by enabling new VMs, \techname{} allocates required cache sizes to the VMs (Fig. \ref{fig:stack-diagram-size1}). Between intervals 10 to 29 (where we run 32 concurrent VMs), the total required cache space by the VMs is greater than the total cache space. In this case, \techname{} reduces the allocated cache sizes, affecting the performance of the VMs (Fig. \ref{fig:Hit-Ratio-32-VM1}). 
	As shown in Fig. \ref{fig:Hit-Ratio-32-VM1}, when only VM0 is online, the average hit ratio is equal to 96\%, then by enabling VM1, VM2, and VM3, the average hit ratios for these VMs are: 77\%, 44\%, 29\%, and 37\%. 
	The performance reduction in VM0 is due to reduced locality of accesses by this VM (in intervals 1-6, the available total cache space is greater than VMs demand). This performance behavior is also observed in another experiment when running only this VM with unlimited cache resources.
	By adding new VMs (interval 7), \techname{} reduces the allocated cache sizes and hence, we experience performance drop in the VMs. In this case, the average hit ratio by VM0 is 28\%.
	Next, \techname{} increases the allocated cache to VM0, since other VMs  demand is low, and hence, cache hit ratio increases to 64\%.

\subsection{Analysis of the Promotion/Eviction Intervals}
\label{sec:analysis}

\begin{figure}[!h]
	\centering
	\includegraphics[scale=.23]{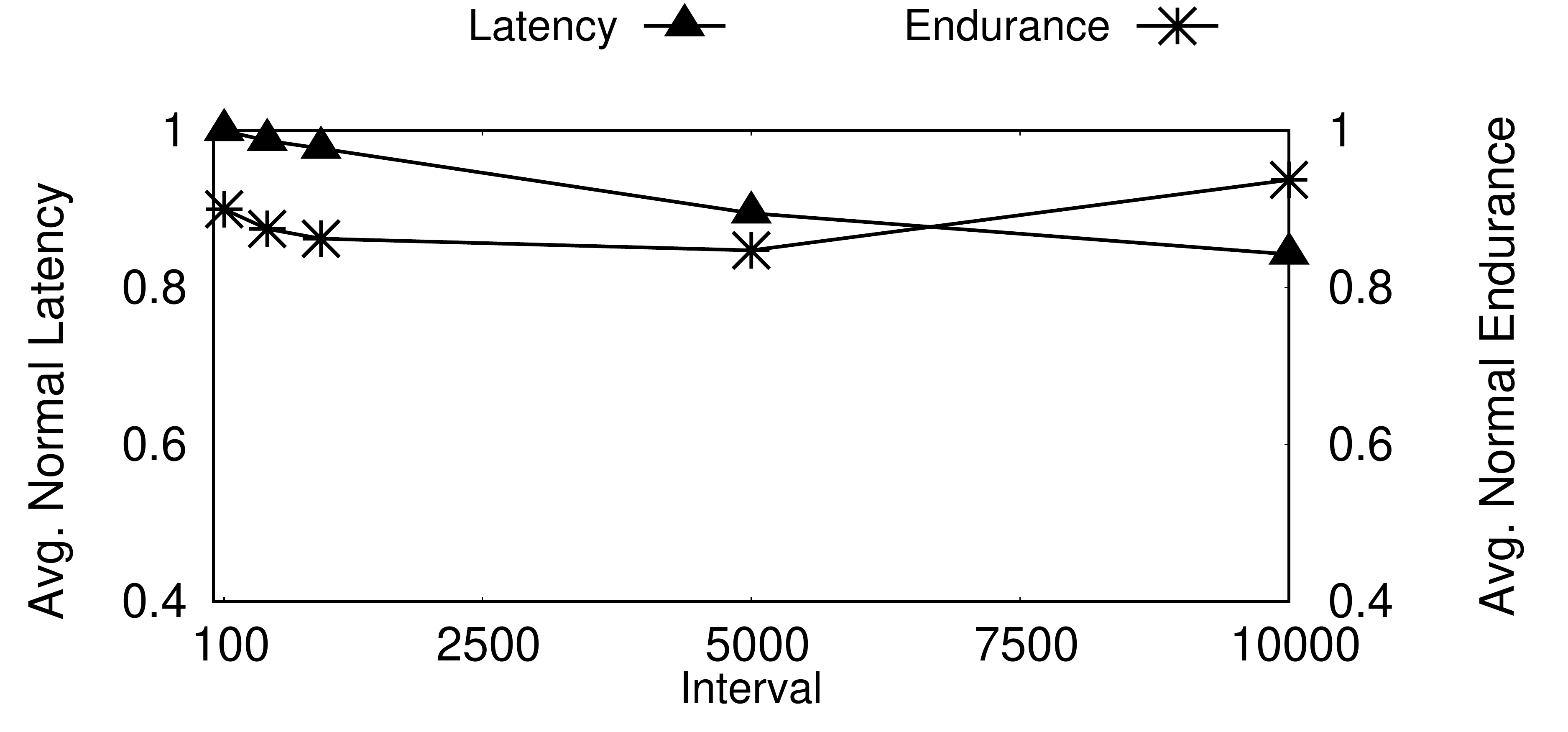}
	\caption{Impact of promotion/eviction intervals on the performance and endurance of \techname{}.}
	\label{fig:intervals}
\end{figure}

The required promotions to ({or} evictions from) SSD are conducted when a fixed number of I/O requests are processed by \techname{}.
Using small intervals enables \techname{} to respond faster to changes in the workload characteristics with the cost of more writes in SSD and therefore, decreasing its lifetime.
To show the impact of interval size on both performance and endurance of \techname{}, we conduct experiments and change the interval {length} from $100$ to $10,000$ {I/O requests processed by the system}.
Fig. \ref{fig:intervals} shows the normalized performance and endurance of \techname{} in various promotion/eviction intervals.
We make four major observations:
\begin{enumerate}[leftmargin=*]
	\item There is a negligible performance improvement between very small promotion/eviction intervals (less than 1,000) and due to the cost of promotion/evictions, employing very small interval sizes is not efficient.
	\item The performance difference between various intervals is about 20\% in all workloads.
	\item In benchmarks with steady workload characteristics throughout their runtime, increasing the interval size does not have any side-effect on the performance.
	Our investigation reveals that this is due to the sparse accesses to the large number of data pages.
	When the intervals are large, such data pages accumulate a high score and therefore, replace the hot data pages in the SSD.
	
\end{enumerate}

We conclude that the selected interval size for promotions/evictions in \techname{} can efficiently balance the performance and endurance in all workloads.

\subsection{\techname{} vs. ECI-Cache}
In this section, we {summarize} how \techname{} overcomes the shortcomings of the latest state-of-the-art I/O caching scheme, ECI-Cache, in terms of cache size, performance, and endurance.

\textbf{Efficient Cache Size Estimation.} {\techname{} employs \URDP{} to estimate the cache size, which results in reduced allocated cache sizes to the VMs. Unlike URD, \URDP{} considers cache write policy in reuse distance estimation, {and thus it} does {\emph{not}} reserve cache blocks for accesses that would not be served by {the} cache.

\textbf{Higher Performance.} {\techname{}, compared to ECI-Cache, improves performance in two ways: 1) {it} {employs} high-performance DRAM in {the} two-level I/O cache structure and 2) {it} {promotes} ({evicts}) only popular (unpopular) data blocks in (from) the I/O cache. 

\textbf{Higher SSD Endurance.}
The proposed per-level write policy management scheme in \techname{} effectively reduces the number of write accesses on the SSD level while {improving the} performance and {preserving} reliability of write-pending data blocks.
In contrast, ECI-Cache buffers both read and write accesses {in} the SSD cache{, which causes more writes into} the SSD cache.

\section{Conclusion}
\label{sec:conclusion}
In this paper, we presented \techname{}, a {new} two-level I/O caching scheme for virtualized platforms.
\techname{} takes advantage of both DRAM and SSD in the I/O caching layer and improves cost, performance, and endurance while preserving the reliability of I/O requests.
The write policy of {the} first {caching} level (i.e., DRAM) is set to RO {(to preserve the storage reliability in the presence of volatile DRAM)} while we use {the} WBWO policy {(to improve SSD‌ endurance)} in the second {caching} level (i.e., SSD). DRAM cache enhances the performance of read requests by buffering read misses while write requests are buffered by the SSD in the second level, providing both {high} performance and  {high} reliability for the write-pending requests. 
\techname{} {further} improves the I/O performance of the running workloads by detecting and buffering \emph{popular} data blocks where  the data blocks {are} not evicted from the cache until they become \emph{unpopular}. 
\techname{} improves {the} endurance of the SSDs by 1) {not buffering} read requests in the SSD cache, which results in {a} significantly reduced number of writes {into} the SSD cache and 2) only promoting \emph{popular} data blocks into the cache and eliminating {the} update {of the SSD}  in {the} case of read misses.
\techname{} improves {the} cost of the I/O cache by 1) allocating a smaller cache size to {each} VM and 2) assigning {an} effective write {policy} to {each} cache level and hence, eliminating the need for using high-cost peripherals such as battery backup to maintain the reliability of write-pending data blocks in {the} case of power {or system failures}.
{\techname{} employs our new} \emph{\URDPfull{}} (\URDP{}) metric, which considers both 1) \emph{request type} and 2) \emph{cache write policy} in cache size estimation and hence, estimates a much smaller cache size {for each} VM while preserving the I/O performance {of the VMs}.
The results of {our real system} experiments {show} that \techname{} provides 45\% and 33.8\% higher performance and endurance and also 51.7\% reduced cache size compared to {the best} state-of-the-art I/O caching {policy} \cite{ahmadian2018eci} in virtualized platforms.



\bibliographystyle{IEEEtran}
\bibliography{IEEEabrv,ref}

\newpage
  \appendices

\IEEEdisplaynontitleabstractindextext

%
\IEEEpeerreviewmaketitle



\ifCLASSOPTIONcaptionsoff
  \newpage
\fi

\end{document}